\documentclass[12pt, referee]{aastex}
\usepackage[dvips]{color}
\newcommand{\re}{\textcolor[rgb]{0,0,0}}
\newcommand{\bl}{\textcolor[rgb]{0,0,0}}

\def\ltsima{$\;\buildrel < \over \sim \;$}
\def\simlt{\lower.5ex \hbox{\ltsima}}
\def\gtsima{$\;\buildrel > \over \sim \;$}
\def\simgt{\lower.5ex \hbox{\gtsima}}

\shorttitle{Water abundance behind interstellar shocks}
\shortauthors{Neufeld et al.}

\begin{document}

\title{The water abundance behind interstellar shocks: results from {\it Herschel}/PACS and {\it Spitzer}/IRS 
observations of H$_2$O, CO, and H$_2$}

\author{David A.~Neufeld\altaffilmark{1}, Antoine Gusdorf\altaffilmark{2}, Rolf G\"usten\altaffilmark{3}, Greg~J.~Herczeg\altaffilmark{4}, Lars Kristensen\altaffilmark{5}, Gary~J.~Melnick\altaffilmark{5}, Brunella Nisini\altaffilmark{6}, Volker Ossenkopf\altaffilmark{7}, Mario Tafalla\altaffilmark{8}, and Ewine~F.~van~Dishoeck\altaffilmark{9,10}}

\altaffiltext{1}{Department\ of Physics \& Astronomy, Johns Hopkins University\,
3400~North~Charles~Street, Baltimore, MD 21218, USA}

\altaffiltext{2}{LERMA, UMR 8112 du CNRS, Observatoire de Paris, \'Ecole Normale Sup\'erieure, 24 rue Lhomond, 75005, Paris, France}

\altaffiltext{3}{Max-Planck Institut f\"ur Radioastronomie, Auf dem H\"ugel 69, 53121 Bonn, Germany}

\altaffiltext{4}{Kavli Institute for Astronomy and Astrophysics, Peking University, Yi He Yuan Lu 5, Hai Dian Qu, 100871 Beijing, P.R. China}

\altaffiltext{5}{Harvard-Smithsonian Center for Astrophysics, 60 Garden Street, Cambridge, MA 02138, USA}

\altaffiltext{6}{Osservatorio Astronomico di Roma, via di Frascati 33, 00040 Monteporzio Catone, Italy}

\altaffiltext{7}{Physikalisches Institut der Universit\"at zu K\"oln, Z\"ulpicher Strasse 77, 50937 K\"oln, Germany}

\altaffiltext{8}{Observatorio Astron\'omico Nacional (IGN), Alfonso XII 3, E-28014 Madrid, Spain}

\altaffiltext{9}{Leiden Observatory, Leiden University, PO Box 9513, 2300 RA Leiden, The Netherlands}

\altaffiltext{10}{Max Planck Institut f\"ur Extraterrestrische Physik, Giessenbachstrasse 1, 85748 Garching, Germany}
\begin{abstract}

\re{We have investigated the water abundance in shock-heated molecular gas, making use of {\it Herschel} measurements of far-infrared CO and H$_2$O line emissions in combination with {\it Spitzer} measurements of mid-IR H$_2$ rotational emissions.  We present far-infrared line spectra obtained with {\it Herschel}'s PACS instrument in range spectroscopy mode towards two positions in the protostellar outflow NGC 2071 and one position each in the supernova remnants W28 and 3C391.  These spectra provide unequivocal detections, at one or more positions, of 12 rotational lines of water, 14 rotational lines of CO, 8 rotational lines of OH (4 lambda doublets), and 7 fine-structure transitions \bl{of atoms or atomic ions. We first used a simultaneous fit to the CO line fluxes, along with H$_2$ rotational line fluxes measured previously by {\it Spitzer}, to constrain the temperature and density distribution within the emitting gas; and we then investigated the water abundances implied by the observed H$_2$O line fluxes.  The water line fluxes are in acceptable agreement with standard theoretical models for nondissociative shocks that predict the complete vaporization of grain mantles in shocks of velocity $v \sim 25$~km/s, behind which the characteristic gas temperature is $\sim 1300$~K and the H$_2$O/CO ratio is 1.2} }

\end{abstract}


\section{Introduction}

{As discussed in detail in the recent review of van Dishoeck et al.\ (2013),} water is a ubiquitous astrophysical molecule that has been observed extensively in comets, circumstellar envelopes around evolved stars, protostellar disks, circumnuclear disks in active galaxies, and interstellar gas clouds.  Astrophysical objects in which water has been detected range in distance from solar system objects to galaxies at redshifts greater than 5.  In the interstellar medium, water vapor can play an important role as a coolant, and its observed abundance is expected to serve as an important astrophysical probe.  Over the past two decades, wide variations in the interstellar water abundance have been inferred from observations of non-masing water transitions, performed with the {\it Infrared Space Observatory (ISO)}, the {\it Submillimeter Wave Astronomy Satellite (SWAS)}, the {\it Odin} satellite, the {\it Spitzer Space Telescope}, and the {\it Herschel Space Observatory}.    In particular, enhanced abundances of water vapor have been reported in the cooling gas behind interstellar shock waves, where water vapor is expected to be the dominant reservoir of gas-phase oxygen, due to the release of ice mantles by sputtering and/or the production of water vapor through endothermic gas phase reactions. {Previous} observational determinations of the water abundances in the {shocked} ISM have been {summarized} by van Dishoeck et al.\ (2013; their Table 4).

Whilst the abundances of water vapor behind interstellar shock waves are -- without question -- several orders of magnitude larger than those present in cold molecular gas clouds, the precision of detailed quantitative measurements of the water abundance have been limited by uncertainties in the physical conditions within the emitting gas.  For example, in their study of water transitions observed with \re{the} Infrared Spectrograph (IRS) on {\it Spitzer} toward the protostellar outflow associated with NGC 2071, Melnick et al.\ (2008; hereafter M08) combined mid-IR observations of pure rotational emissions from H$_2$ and H$_2$O to infer a water abundance H$_2$O/H$_2$ in the range $2 \times 10^{-5}$ and $6 \times 10^{-4}$ (the upper end of which is broadly consistent with theoretical predictions).  The factor of 30 uncertainty in the inferred range largely reflected uncertainties in the density of the emitting gas; because of the large spontaneous radiative rate for mid-IR H$_2$O transitions, the relevant rotational states are subthermally populated (unlike those of H$_2$); thus the inferred H$_2$O abundance is \re{roughly} inversely proportional to the assumed gas density.  M08 suggested that future observations of high-lying rotational states of CO -- which could be performed with the {then}-upcoming launch of {\it Herschel} -- could help place stronger constraints upon the water abundance.

In this paper, we present new observations of far-IR H$_2$O and CO emissions from shocked interstellar gas, carried out with the Photodetector Array Camera and Spectrometer (PACS; Poglitsch et al.\ 2010) instrument on {\it Herschel} (Pilbratt et al.\ 2010) toward three sources: the protostellar outflow associated with NGC 2071 (in which two positions were observed), and the supernova remnants 3C391 and W28.   These sources represent archetypal examples of interstellar shock waves caused by the interaction of a protostellar outflow or a supernova remnant with an ambient molecular cloud, and have been discussed in recent studies of line emissions observed with {\it Spitzer} (M08; Neufeld et al.\ 2007, hereafter N07).  To recap these discussions briefly,  NGC 2071 lies at a distance of 390 pc, (Anthony-Twarog 1982) within the Orion B molecular cloud complex, and is an \re{active} region of \re{intermediate-mass} star formation; the positions observed in this study lie within a prominent bipolar outflow on either side of a central cluster of infrared sources, at projected distances of 0.13 and 0.28 pc from the infrared source IRS1.  3C391 is an X-ray and radio-bright supernova remnant lying at a distance of 9 kpc within the Galactic plane; in this study we have targeted the "broad molecular line'' region (Reach \& Rho 1999), which exhibits prominent 1720 MHz OH maser emission along with broad (FWHM $\sim 20$~km~s$^{-1}$) non-masing emission lines of CO, CS and HCO$^+$.  W28 is a mixed-morphology supernova remnant, located at a distance of $\sim 1.9$~kpc, that combines a shell-like radio structure with a centrally filled diffuse X-ray emission structure; as in the case of 3C391, we have targeted a region of prominent OH maser emission in the 1720 MHz (satellite) line, a phenomenon that is believed to be a tracer of the interaction of supernova remnants with molecular clouds (e.g. Wardle \& Yusef-Zadeh 2002).  \bl{All three sources show broad molecular line emission, with full widths at zero intensity in the range 40 to 50~km~s$^{-1}$, indicating the presence of supersonic motions that can give rise to shock waves (see, for example, Neufeld et al.\ 2000 for NGC 2071; Frail \& Mitchell 1998 for 3C391; and Gusdorf et al.\ 2012 for W28).}

In Section 2 below, we describe the observations and data reduction, and we present the far-infrared spectra thereby obtained.  In Section 3, we discuss how the far-infrared line fluxes measured with {\it Herschel}/PACS can be combined with previous {\it Spitzer} observations of H$_2$ pure rotational transitions to obtain estimates of the physical conditions in the emitting gas and the water abundance.  In Section 4, these water abundance estimates are discussed in the context of astrochemical models.  A brief summary follows in Section 5.

\section{Observations, data reduction and observed far-infrared spectra}

The {\it Herschel}/PACS data presented here were all acquired in range spectroscopy mode.  The coordinates of the observed positions, observing dates, integration times, and Astronomical Observing Request (AOR) numbers are given in Table 1.  The positions of the PACS observations are indicated in Figures 1 -- 3, which superpose the PACS ``spaxel" {(spatial pixel)} positions on maps of each source.  In the case of NGC 2071, the background image is a {\it Spitzer}/IRAC map, whilst for 3C391 and W28 it is a radio continuum image.  The PACS spectrometer obtains spectra at  25 spaxels of size 
$9^{\prime \prime}.4  \times  9^{\prime \prime}.4$  arranged in a roughly rectangular $5 \times 5$ grid.

The observations of NGC 2071 were performed as part of the ``Water in Star-forming Regions \re{with Herschel}" (WISH) program (PI E.~van Dishoeck) with spectral scans covering the 70 -- 105 micron and 102 -- 200 micron regions in the second and first orders of the PACS spectrometer.  For each of 2 positions in NGC 2071, the data were obtained with 2 repetitions performed in chopping/nodding mode with a chopper throw of 3 arcmin and Nyquist spectral sampling.  The observations of 3C391 and W28 were performed as part of the ``Warm and Dense ISM" (WADI) program (PI V.~Ossenkopf) using the so-called ``SED B2B plus long R1" and the ``SED B2A plus short R1" modes of the spectrometer.  The data were obtained with a single repetition performed in chopping/nodding mode with a chopper throw of 6 arcmin and Nyquist spectral sampling.
 
The data were processed using version 8 of the Herschel Interactive Processing Environment (HIPE).  The spectral resolution ranges from 2000--3000 in the 70-100 $\mu$m region and 1000--2000 in the 100--200 $\mu$m region.  The spectra were resampled onto a wavelength grid with $\sim 2$ pixels per resolution element.  Fluxes were calculated by comparing the flux to the telescope background and using Neptune as a calibrator.  {The accuracy of the absolute flux calibration of the PACS instrument has been determined previously from observations of thirty standard calibrators\footnote{See section 2.2.1 in ``PACS Spectroscopy performance and calibration,'' PACS/ICC document ID PICC-KL-TN-041, 16 June 2011, prepared by B.\ Vandenbussche}.  The r.m.s.\ errors derived from these observations were $11 - 12 \%$ and the peak-to-peak errors were $\pm 30\%$.}  

Additional processing was carried out using custom routines developed in the Interactive Data Language (IDL).  These were used to fit continua to line-free regions of the spectra, and to fit Gaussian profiles to the detected spectral lines, using the Levenberg-Marquardt algorithm. In Figures 4 -- 7, we present the continuum-subtracted spectra obtained toward each source.  These spectra are sums over the 25 PACS spaxels.  Vertical markings indicate the wavelengths of possible spectral lines within the PACS bandpass, with CO, H$_2$O, and OH transitions marked in blue, red, and green respectively, and fine structure transitions of atoms and atomic ions marked in magenta.  
Single vertical marks indicate lines that were securely detected, at good signal-to-noise ratio ($\simgt$5), while pairs of shorter marks indicate the positions of lines that were not detected, detected at low signal-to-noise ratio, or showed an unusual appearance that cast doubt upon the reality of the feature.  A complete list of all marked transitions is given in Table 2, which lists the fluxes measured for good signal-to-noise ratio lines.  \re{Here, we focus on the CO and H$_2$O line emissions observed with PACS; these emissions originate primarily in molecular gas that has been heated by nondissociative shocks, and are excited primarily by collisional excitation with H$_2$.   In the present paper, we will not consider further the fine structure emissions observed from atoms and atomic ions, which are primarily excited in faster, dissociative shocks, nor the rotational emissions from OH, for which radiative excitation typically plays a critical role (e.g. van Dishoeck et al.\ 1987, Melnick et al.\ 1990).} 

The flux values given \re{ in Table 2} for CO and H$_2$O were used in the excitation analysis of Section 4.  In addition, we list the fluxes for 7 pure rotational transitions of H$_2$, observed previously with {\it Spitzer}/IRS (N07, Neufeld et al.\ 2009), and -- in the case of NGC 2071 NE -- for four water transitions observed with {\it Spitzer} (M08).  Finally, for 3C391 and W28, we use {\it ISO} measurements (Reach \& Rho 2000) of the H$_2$ S(3)/S(9) ratio to estimate the H$_2$ S(9) flux.  \re{The fluxes given in Table 2 and used in the excitation analysis are sums over the 25 PACS spaxels}\footnote{\bl{Because the resolution of PACS is wavelength-dependent at wavelengths above 100~$\mu$m (Poglitch et al.\ 2010), with a full-width at half maximum that is comparable to -- or larger than -- the spaxel size, and because there are small offsets between the spaxel positions for the blue and red SED modes, our use of summed fluxes yields line ratios that are more robust than those that could be obtained from the brightest few pixels alone.  However, our choice of using the total fluxes, summed over 25 spaxels, does entail a cost in the resultant signal-to-noise ratio (Karska et al.\ 2013; their Appendix B)}}

In addition to the total line fluxes measured by PACS, we have also determined the distribution of line fluxes amongst the 25 spaxels.  These distributions are shown in Figures 8 - 11 for 4 selected lines observed with PACS, \re{along with the mid-infrared H$_2$~S(5) line observed by {\it Spitzer} (N07).  In the case of the H$_2$ S(5) line, the {\it Spitzer} spectral line maps are shown both at their native resolution $\sim 2^{\prime\prime}$ (top left panel) and rebinned onto the footprint of the {\it Herschel}/PACS spaxels (top right panel).}  Clearly, in any given source, all five line emissions show a similar spatial distribution, suggesting that they originate in the same material.

\section{Derivation of the physical parameters and water abundance}

In modeling the excitation of CO, H$_2$ and H$_2$O, we have solved the equations of statistical equilibrium using a code described previously by Neufeld \& Kaufman (1993). 
Here, an escape probability method was used to treat the effects of radiative trapping in optically-thick lines; such effects are most important for H$_2$O and entirely unimportant for H$_2$.  We adopted the rotational energies given by Nolt et al.\ (1987), Dabrowski (1984) and Kyro (1981) espectively for CO, H$_2$ and H$_2$O; and the spontaneous radiative transition probabilities given respectively by Goorvitch (1994), Wolniewicz et al.\ (2008) and Coudert et al.\ (2008).  Assuming H$_2$ to be the dominant collision partner, we made use of the state-to-state rate coefficients given by Yang et al.\ (2010), Flower (1998), and Daniel et al.\ (2011) for the collisional excitation of CO, H$_2$ and H$_2$O by H$_2$.  \re{ Except in regions very close to strong sources of infrared continuum radiation (such as the circumstellar envelopes of evolved stars), radiative pumping is unimportant in the excitation of these molecules and was neglected in our analysis.}

The H$_2$ and CO rotational line strengths listed in Table 2 are conveniently represented on ``rotational diagrams", in which the column density per magnetic substate -- computed under the assumption that the optical depth is negligible -- is plotted against the energy of the upper state on a log-linear scale.  For gas in local thermodynamic equilibrium (LTE) at a single temperature, the resultant rotational diagrams yield a straight line with a slope inversely proportional to the gas temperature.   As is typically found for other sources observed with {\it Herschel}/PACS (e.g. \bl{Herczeg et al.\ 2012, Goicoechea et al.\ 2012, Manoj et al.\ 2013, Green et al.\ 2013}), the rotational diagrams for CO (Figures 12 -- 15, upper right panels, \re{blue symbols}) all exhibit positive curvature, indicating either (1) the presence of a range of gas temperatures along the sight-line and within the beam; and/or (2) the presence of subthermal excitation.  For the case of CO, Neufeld (2012) showed that condition (2) above could be sufficient to yield moderate positive curvature even for an isothermal emitting region.  In the case of H$_2$, however, the sources observed here also show positively-curved rotational diagrams \re{(Figures 12 -- 15, upper right panels, red symbols)}, and these {\it cannot} be accounted for by an isothermal emitting region unless unreasonably low densities ($n_{\rm H} \simlt 10\rm \, cm^{-3}$ ) are posited.  In addition, as noted in Neufeld et al.\ (2006) for the sources HH54 and HH7, the rotational diagrams show a zig-zag behavior that implies a non-equilibrium ortho-to-para ratio (OPR) for H$_2$; evidently the emitting gas has not been warm long enough to acquire the OPR ($\sim 3$) that would obtain in equilibrium at its current temperature, but instead retains a ``memory" of an earlier epoch in which it was much cooler.  The departures from an equilibrium OPR value are more pronounced for low-lying rotational states, a behavior that suggests the rate of equilibration to be an increasing function of temperature.  This is readily understood if reactive collisions with H are the dominant \re{process for ortho-para conversion} \bl{in warm gas}, as they possess an activation energy barrier $\sim 3900$~K (Schofield 1967).  

Motivated by the positive curvature in the H$_2$ rotational diagrams, we followed previous work (e.g. Neufeld et al.\ 2009, hereafter N09) in assuming a power-law distribution of gas temperatures, with the amount of material at temperature between $T$ and $T + dT$ assumed proportional to $T^{-b} dT$ over a wide range of temperatures (100 K to 5000 K).  We obtained estimates for the gas density, $n({\rm H}_2)$, and the power-law index, $b$, from a best fit to the {\it Spitzer} H$_2$ S(0) -- S(5) fluxes and the PACS CO fluxes tabulated in Table 2.  We did not include the H$_2$ S(7) or S(9) line fluxes, because those lines trace considerably hotter gas than that traced by the CO and H$_2$O lines that we have detected with PACS.   In this analysis, we made the following assumptions: (1) a CO abundance corresponding to the gas-phase elemental abundance of carbon in diffuse molecular clouds (Sofia et al.\ 2004), $n({\rm CO})/n({\rm H}_2) = 2\,n_{\rm C}/n_{\rm H} = 3.2 \times 10^{-4}$; (2) an H$_2$ OPR with the temperature dependence given by N09 (their eqn 1), in which the initial OPR ratio, $\rm OPR_0$ and the time period for equilibration, $\tau$ are treated as free parameters; (3), for purposes of determining the line optical depths, an $\rm H_2$ column density of $\tilde{N}({\rm H}_2) = 10^{20}$~cm$^{-2}$ per km~s$^{-1}$, the canonical value
 for nondissociative interstellar shocks\footnote{\re{Here, $\tilde{N}({\rm H}_2)$ is the optical depth parameter defined by Neufeld \& Kaufman (1993).  In the large velocity gradient limit, which applies behind nondissociative shock waves, it is simply equal to $n({\rm H}_2)/\vert dv_z/dz \vert$, where $dv_z/dz$ is the velocity gradient in the direction of shock propagation.  Since the $n({\rm H}_2)$ and $dv_z/dz$ are both roughly proportional to the preshock density, $\tilde{N}({\rm H}_2)$ is roughly independent of the preshock density.}} (e.g.\ Neufeld \& Kaufman 1993); and (4) a constant gas density.  The assumption of constant density, in particular, is clearly an idealization.  \re{However, whereas the {\it Spitzer}-detected H$_2$ line emissivities are roughly independent of gas density (the H$_2$ level populations being close to LTE), the emissivities of CO and H$_2$O lines within the PACS wavelength range are roughly proportional to $n({\rm H}_2)$ in the density regime of present interest.  Thus, the H$_2$O/CO abundance ratio inferred from the relative strength of the PACS-detected CO and H$_2$O lines is relatively insensitive to the assumed gas density.}
  
In the upper left panels of Figures 12 -- 15, we present the goodness-of-fit \bl{to the CO and H$_2$ line fluxes alone}, as a function of $n({\rm H}_2)$ and $b$.  The black contours plotted here show the \bl{(unweighted)} r.m.s. value of log$_{10}$(observed/fitted line flux), which we adopt as a measure of the goodness-of-fit.   We favor this measure, rather than $\chi^2$, because the measurement errors are likely dominated by systematic rather than statistical uncertainties.  Thus, the use of $\chi^2$ as a statistic tends to over weight those lines for which the signal-to-noise ratio is largest.  The red contours show the beam-averaged H$_2$ column density, $N({\rm H}_2)$, needed to fit the absolute fluxes, and are labeled with log$_{10}[N({\rm H}_2)$ in cm$^{-2}$].  In the upper right panels of Figures 12 -- 15, we show the H$_2$ and CO rotational diagrams obtained for the best-fit parameters (black lines).

In modeling the excitation of water, we consider two free parameters: the water abundance relative to CO, $n({\rm H_2O})/n({\rm CO})$, and the minimum temperature at which water is present, $T_{\rm w}$.  Here, motivated by astrochemical models that posit a dramatic increase in the water abundance at gas temperatures $\simgt 400$~K \bl{(due to high temperature chemistry; see Kaufman \& Neufeld 1996, hereafter KN96)} or for shock velocities $\simgt 25 \, \rm km \, s^{-1}$ behind which the gas temperature is $\simgt 1300$~K \bl{(due to ice sputtering; see Draine et al.\ 1983, hereafter D83)} , we adopt a very simple assumption that the water abundance is zero for $T < T_{\rm w}$ and has some constant value for $T \ge T_{\rm w}$.  Given the best-fit physical parameters determined from a simultaneous fit to the H$_2$ and CO transitions, as described above, we then considered the goodness-of-fit to the H$_2$O line fluxes, as a function of $T_{\rm w}$ and ${\rm log}_{10}[n({\rm H_2O})/n({\rm CO})]$.  Results are shown in the lower left panels of Figures 12 -- 15, where we show the r.m.s. value of log$_{10}$(observed/fitted line flux) for the water line fluxes listed in Table 2 (\bl{including the mid-IR lines observed by {\it Spitzer} toward NGC 2071 NE}).  Here, \bl{an identical filling factor is assumed for all transitions}, an OPR of 3 is assumed for water, and the $n({\rm H_2O})/n({\rm CO})$ ratio is allowed to range up to $(n_{\rm O}-n_{\rm C})/n_{\rm C} \sim 1.2$, where $n_{\rm O}/n_{\rm H} = 3.5 \times 10^{-4}$ is the gas-phase abundance of oxygen in diffuse atomic clouds (Cartledge et al.\ 2001).  This maximum value is the $n({\rm H_2O})/n({\rm CO})$ ratio that could be attained if grain mantles were fully sputtered and H$_2$O accounted for all gas-phase oxygen not bound as CO.  {Our computation of $n({\rm H_2O})/n({\rm CO})$ and $T_{\rm w}$ was performed assuming the best-fit parameters for the gas density, $n({\rm H}_2)$, and the power-law index, $b$, derived previously.  Thus, rather than obtaining a simultaneous fit to all the data within a four-dimensional parameter space, we perform two sequential fits within the two-dimensional parameter spaces defined by $n({\rm H}_2)$ and $b$, and by $n({\rm H_2O})/n({\rm CO})$ and $T_{\rm w}$.  This simplification is justified by the fact that the primary degeneracies are between $n({\rm H}_2)$ and $b$, and between $n({\rm H_2O})/n({\rm CO})$ and $T_{\rm w}$.  Within the range of acceptable parameters for $b$, for example, the derived value for $T_w$ shows a much weaker dependence on $b$ than it does upon $n({\rm H_2O})/n({\rm CO})$.}

Because H$_2$O is an asymmetric top, rather than a diatomic molecule, its excitation does not naturally lend itself to being representing by a rotational diagram.  Thus, in the lower right panels of Figures 12 -- 15, we simply present the best-fit log$_{10}$(observed/fitted line fluxes) for all transitions of H$_2$O (along with those for H$_2$ and CO).  Here, the transitions are ordered vertically by energy of the upper state.  In Table 3, we list best-fit values for the six parameters we adjusted to obtain this fit.  \re{ In the case of all four observed positions, the lower right panels of Figures 12 -- 15 show no evidence for para-H$_2$O transitions (green) being systematically overpredicted or underpredicted relative to ortho-H$_2$O transitions (red); thus, the data appear broadly consistent with the assumed OPR of 3 for water.}

\section{Discussion}

\subsection{Physical conditions}

The best-fit densities and power-law indices inferred for the supernova remnants W28 and 3C391 all lie within previous estimates, reported by Yuan and Neufeld (2011, hereafter Y11), that were based on an earlier dataset that did not include the PACS CO line fluxes.  \bl{In the case of W28, the best-fit density, $n({\rm H}_2) \sim 4000\rm \, cm^{-3}$, is broadly consistent with a previous study by Gusdorf et al.\ (2012).  In that study, the strengths of CO lines up to $J=11-10$ were compared with shock models computed for preshock densities $n({\rm H}_2) =$ 500, 5000, $5 \times 10^{4}$ and $5 \times 10^5\rm \, cm^{-3}$; the best fit was obtained for preshock density $n({\rm H}_2) =5000  \rm \, cm^{-3}$, which corresponds to a typical density of $\sim 10^4 \rm \, cm^{-3}$ in the postshock emitting region.}
The parameters inferred for NGC 2071 NE lie within the (broad) range inferred by M08 on the basis of H$_2$ line fluxes alone, \bl{although the inferred densities and power-law indices are considerably smaller than those derived by Giannini et al.\ (2011; hereafter G11).  This discrepancy likely results from the inclusion of ground-based measurements of H$_2$ vibrational emissions (along with {\it Spitzer} measurements of H$_2$ pure rotational lines) in the analysis performed by G11.  The fact that these two analyses disagree may suggest a shortcoming in our assumption of a constant gas density, as suggested by Yuan (2012).}

Overall, the best-fit power-law indices lie in a narrow range (2.7 to 3.2) that appears to be typical of interstellar shock waves (e.g. N09, Y11, Yuan 2012), \bl{at least as inferred from observations of H$_2$ rotational emissions}.  \bl{Given typical interstellar magnetic fields, planar shocks of a single velocity yield H$_2$ rotational diagrams that are nearly flat, the bulk of the H$_2$ emission occurring within a postshock region of roughly constant temperature.  Thus, the curved rotational diagrams that we have observed suggest a distribution of shock velocities.}  \re{Such a distribution might plausibly arise in unresolved bow shocks that present a range of shock velocities along a given sightline and within a given beam; as indicated by the H$_2$ line maps plotted in Figures 8 - 11, the shock-excited line emission clearly exhibits structure that is unresolved at the resolution of PACS.}
As discussed in Y11, the inferred values for $b$ are somewhat smaller than that expected for paraboloidal bow shocks, $b=3.8$, but might be understood for bow shocks with a characteristic shape that is somewhat flatter than a paraboloid (i.e.\ which presents a somewhat larger surface area at large angles to its axis).  

{Given the best-fit densities and power-law indices determined from the CO and H$_2$ line ratios, the absolute line fluxes indicate warm ($T> 100$~K) H$_2$ column densities in the range $0.3 - 3 \times 10^{21}\,\rm cm^{-2}$, averaged over the regions observed with PACS (see red contours in the upper left panels of Figures 12 - 15).  These values are typical of the column densities expected behind non-dissociative molecular shocks (e.g.\ Neufeld et al.\ 2006).}

\subsection{Water abundance}

In standard models for the chemistry of nondissociative shock waves\footnote{\re{Given typical interstellar magnetic fields, shocks slower than $\sim$~50~km~s$^{-1}$  are nondissociative, whereas shocks faster than $\sim$~50~km~s$^{-1}$ are dissociative and lead to the destruction of H$_2$, CO, and other molecules by collisional dissociation.}} (e.g.\ D83, KN96, \bl{
Gusdorf et al.\ 2008, Flower \& Pineau des For\^ets 2010}), 
shocks faster than a threshold speed yield postshock gas temperatures sufficient to produce water rapidly through a pair of neutral-neutral reactions with activation energy barriers: $\rm O(H_2,H)OH(H_2,H)H_2O$.  For typical interstellar magnetic fields, $B \sim (n_{\rm H}/{\rm cm}^{-3})^{1/2}\,\mu \rm G$, and in the absence of strong sources of ionizing \bl{and dissociating} radiation, the threshold shock speed for rapid conversion of O to H$_2$O is $\sim 15$~km~s$^{-1}$, sufficient to yield postshock gas temperatures in excess of $\sim 400$~K.  
However, this threshold may not be entirely relevant, since -- as implied by the small O$_2$ and H$_2$O abundances measured in dense cold molecular clouds (Bergin et al.\ 2000) -- gas-phase atomic oxygen may not be a major constituent of the preshock gas.  Instead, the dominant reservoir of oxygen may be \bl{grain mantles composed of ice and other oxygen-containing materials}.  In that case, the relevant process is the sputtering of \bl{grain mantles}.  The fraction of the material released to the gas phase is expected to be a strong function of velocity: for a 0.3 micron diameter grain composed of pure water ice, D83 predicted a release fraction that increases from $\sim 10\%$ behind a 20 km/s shock to $\sim 100\%$ behind a 25~km~s$^{-1}$ shock\footnote{\re{ One limitation of the D83 results is that they are computed for a single grain size near the upper end of the expected size distribution.  Given a distribution of grain sizes, the transition between a small released ice fraction and complete mantle erosion will be somewhat more gradual.}}.  On the other hand, Flower \& Pineau des For\^ets (1994; hereafter FP94) obtained a considerably smaller threshold velocity for the release of icy grain mantles, as small as 10~km~s$^{-1}$.  However, as noted by Draine (1995),  FP94 assumed  threshold energy for grain sputtering that was too small by a factor of 4 (at least for pure water ice).  Results similar to those of D83 have been obtained by Jim\'enez-Serra et al.\ (2008; see their Figure 7) and by \bl{Gusdorf et al.\ (2008)}.
\bl{In these models, the oxygen released from grain mantles quickly ends up in the form of water vapor, regardless of whether the sputtering initially releases O, OH, or H$_2$O, because the postshock temperatures are high enough that O and OH are rapidly converted to H$_2$O by reaction with H$_2$.} 

The characteristic gas temperature behind a nondissociative C-type shock of speed, $v_s$, can be approximated by $T_s = 375 b_M^{-0.36} [v_s/10 \rm \,km\,s^{-1}]^{1.35}$ (Neufeld et al.\ 2006; based upon detailed modeling by NK93), where $b_M = B/(n_{\rm H}/{\rm cm}^{-3})^{1/2}\,\mu \rm G.$   If, following D83, we adopt 25~km~s$^{-1}$ as the threshold velocity for the complete sputtering of grain mantles, the corresponding threshold gas temperature is 1300~K, given standard interstellar magnetic fields ($b_M = 1$).  Above this threshold, H$_2$O/CO ratios $\sim (n_{\rm O}-n_{\rm C})/n_{\rm C} \sim 1.2$ are expected.  Thus, we take $T_{\rm w}=1300$~K, $n({\rm H_2O})/n({\rm CO}) = 1.2$ as the standard theoretical predictions against which the results obtained in Section 3 are to be compared.  These theoretical predictions are denoted by the red triangles in the lower right panels of Figures 12 -- 15.  

\bl{Several additional considerations might modify the ``standard predictions" represented by the red triangles in Figure 12 -- 15.  First, the predicted $T_{\rm w}$ values apply for typical interstellar magnetic fields ($b_M=1$); larger fields would move the red triangles to the left, and smaller fields to the right.  Second, they assume the ionization fraction appropriate to molecular gas that is well-shielded from ultraviolet radiation.   If a local source of ultraviolet radiation is present, the ion-neutral coupling length can be reduced, leading to a larger gas temperature for a given shock velocity: this would move the red triangles to the right.  The supernova remnant 3C391 exhibits fine structure emissions from [OIII], suggesting that fast dissociative shocks are present along with the slower nondissociative shocks that produce the observed H$_2$ and CO emissions.  These dissociative shocks will be sources of UV radiation, although the degree to which the nondissociative molecular shocks are irradiated will depend strongly upon the geometry.  Third, if a significant ultraviolet radiation field is indeed present at the location of the warm molecular gas, photodissociation can limit the water abundance behind the shock front; however, at a temperature of $1300$~K or greater, neutral-neutral reactions are extremely efficient in reforming H$_2$O.  Thus, the effects of photodissociation are only likely to be significant in downstream material that has already cooled; in this region, elevated OH abundances may be present as a result of H$_2$O photodissociation.  Fourth, water ice does not appear to be sufficiently abundant to account for the oxygen depletion in dense molecular clouds (e.g.\ Jenkins 2009; Whittet 2010).  Thus, other oxygen-bearing molecules within the grain mantle presumably account for a significant fraction of the oxygen budget.  These molecules, as yet undetected by infrared spectroscopy of grain mantles, have been referred to as Unidentified Depleted Oxygen (UDO; Whittet 2010); depending upon their binding energy, they might have different sputtering thresholds than water ice.}

Toward W28, the best-fit values for $T_{\rm w}$ and $n({\rm H_2O})/n({\rm CO})$ are evidently in good agreement with the standard theoretical predictions.  Toward the other positions, 
the {\it best-fit} H$_2$O/CO ratios are significantly smaller (factor 4 to 16) than the standard theoretical predictions, as are the temperatures (factor of 3 to 8).  However, in all sources there is a significant degeneracy between $n({\rm H_2O})/n({\rm CO})$ and $T_{\rm w}$; this is clear in Figures 12 -- 15 (lower left panel), where the plotted contours define a ``valley", extending from the lower left to the upper right of the panel, within which a good fit to the data is found.  \re{Not surprisingly, if the assumed value of $T_{\rm w}$ is increased above its best-fit value, a corresponding increase in the assumed water abundance above $T_{\rm w}$ tends to compensate for that increase.}  Thus, in all sources, the standard theoretical predictions for $n({\rm H_2O})/n({\rm CO})$ and $T_{\rm w}$ appear to be in acceptable agreement with the data, \bl{notwithstanding the additional considerations described in the previous paragraph}.  If we now assume, as a {\it prior}, that the $n({\rm H_2O})/n({\rm CO})$ ratio is 1.2 above the threshold temperature, $T_{\rm w}$, we obtain best-fit $T_{\rm w}$ values in the range 1000 -- 1600~K (see Table 3): these values lie within a factor 1.3 of the standard theoretical prediction, $T_{\rm w} = 1300$~K.  {This result suggests that the interstellar oxygen not bound as CO resides primarily in grain mantles, in the form of ice and other materials with a similar threshold for sputtering, and that this oxygen is converted to water vapor by gas-phase reactions, once released from the grain mantle.}

\section{Summary}

\noindent\re{1.  We have used {\it Herschel}'s PACS instrument in range spectroscopy mode to obtain far-infrared line spectra obtained towards two positions in the protostellar outflow NGC 2071 and one position each in the supernova remnants W28 and 3C391.    We obtained unequivocal detections, at one or more positions, of 41 spectra lines or blends, comprising 12 rotational lines of water, 14 rotational lines of CO, 8 rotational lines of OH (4 lambda doublets), and 7 fine-structure transitions of C$^+$, N$^+$, O, O$^{2+}$, and P$^+$.} 

\noindent\re{2.  In combination with previously-reported {\it Spitzer} measurements of the H$_2$ S(0) - S(5) line fluxes towards these four positions, along with previous {\it Spitzer} detections of mid-infrared H$_2$O emissions, we have obtained a simultaneous fit to the H$_2$, CO and H$_2$O line fluxes.  The positively-curved H$_2$ rotational diagrams obtained for these sources imply that a range of gas temperatures is present along
a given sightline and within a given spaxel.  For an assumed CO/H$_2$ abundance ratio of $3.2 \times 10^{-4}$, a simultaneous fit to the CO and H$_2$ rotational diagrams can be obtained for H$_2$ densities ranging from $\sim 4000 $ to $\sim 10^5\,\rm cm^{-3}$ for the four observed positions.  For an assumed powerlaw distribution of gas temperatures, with the amount of material at temperature $T$ to $T+dT$ assumed proportional to $T^{-b}dT$ for temperatures in the range 100 -- 5000~K, the best fit to the H$_2$ and CO fluxes is obtained for $b$ in the narrow range 2.7 -- 3.2 for the four observed positions.   Such a distribution might plausibly arise in unresolved bow shocks that present a range of shock velocities along a given sightline and within a given beam.}  

\noindent\re{3.  Assuming that the water abundance rises sharply to a constant value at some temperature, $T_{\rm w}$, an assumption motivated by theoretical models for the chemistry behind non-dissociative molecular shock waves, we obtained best-fit values for $T_{\rm w}$ ranging from $\sim 200$ to $\sim 1300$~K at the four observed positions.  Above $T_{\rm w}$, the H$_2$O/CO abundance ratio yielding the best fit to the H$_2$O line fluxes ranged from 0.12 to 1.2,  The water abundance and the value of T$_{\rm w}$ are significantly degenerate, however, and acceptable fits to the data can be obtained with the assumption of a H$_2$O/CO abundance ratio of 1.2 above a threshold temperature in the range $T_{\rm w}=1000$ - 1600 K for the four positions.
This result is consistent with theoretical models that predict the complete vaporization of icy grain mantles in shocks of velocity $v \sim 25$~km/s behind which the characteristic gas temperature is $\sim 1300$~K.  {It suggests that the interstellar oxygen not bound as CO resides primarily in grain mantles, in the form of ice and other materials with a similar threshold for sputtering, and that this oxygen is converted to water vapor by gas-phase reactions, once released from the grain mantle.}}

\acknowledgments

PACS has been developed by a consortium of institutes led by MPE (Germany) and including UVIE (Austria); KU Leuven, CSL, IMEC (Belgium); CEA, LAM (France); MPIA (Germany); INAF-IFSI/OAA/OAP/OAT, LENS, SISSA (Italy); IAC (Spain). This development has been supported by the funding agencies BMVIT (Austria), ESA-PRODEX (Belgium), CEA/CNES (France), DLR (Germany), ASI/INAF (Italy), and CICYT/MCYT (Spain).  D.A.N. gratefully acknowledges the support of Research Support Agreement No.\ 1465490 issued by NASA/JPL.  \bl{A.~G. acknowledges support by grant ANR-09-BLAN-0231-01 from the French Agence Nationale de la Recherche as part of the SCHISM project.}

{}

\begin{deluxetable}{lcccc}
\tablecaption{Observing parameters}
\tablehead{  				&  NGC2071 NE     & NGC2071 SW      & 3C391   & W28      \\}  
\startdata

			RA (J2000)$^a$ 	& $\rm 05h\,47m\,07.80s $ 
							& $\rm 05h\,46m\,57.20s $
							& $\rm 18h\,49m\,22.30s $
							& $\rm 18h\,01m\,52.30s $ \\
		   	Dec (J2000)$^a$ & $\rm  +00^0\,22^\prime\,44^{\prime\prime}.0$ 
		   					& $\rm  +00^0\,20^\prime\,20^{\prime\prime}.1$ 
		   				    & $\rm -00^0\,57^\prime\,22^{\prime\prime}.0$ 
		   					& $\rm -23^0\,19^\prime\,25^{\prime\prime}.0$ \\
		   	Date			& 2011 Sep 11 & 2011 Sep 11 & 2011 Sep 27 & 2011 Apr 05\\
		   	Duration		& 6337~s & 6337~s & 7842~s & 7850~s \\
		   	AORS 			& 1342228464 & 1342228463 & 1342219807 & 1342217942 \\
		        			&            &            & 1342219808 & 1342217944 \\
\enddata
\tablenotetext{a}{Commanded position for central spaxel}
\end{deluxetable} 

\begin{deluxetable}{llccrrrr}
%{|l|l|r|r|r|r|r|}
\tabletypesize{\scriptsize}
\tablewidth{0pt}
\tablecaption{Measured line fluxes$^a$ ($10^{-16}\,\rm W \, m^{-2}$) } 
\tablehead{
Molecule  & Transition   & Wavelength  & $E_U/k$ & NGC 2071  & NGC 2071 & 3C391   & W28      \\
		  &              & (micron)& (K) & NE lobe \phantom{0} & SW lobe \phantom{0}\\ 
& & & & \phantom{NGC 2071}	& \phantom{NGC 2071}	& \phantom{NGC 2071}	& \phantom{NGC 2071}	   }
\startdata

H$_2$ & S(9) &   4.69 & 10262 &   ...   &   ...   &   77.69 &   36.98 \\
H$_2$ & S(7) &   5.51 &  7197 &  112.84 &   75.97 &  217.73 &  108.67 \\
H$_2$ & S(5) &   6.91 &  4586 &  235.45 &  146.86 &  523.71 &  301.85 \\
H$_2$ & S(4) &   8.03 &  3475 &  113.23 &   74.11 &  118.89 &  103.55 \\
H$_2$ & S(3) &   9.66 &  2504 &  123.08 &  113.58 &  335.29 &  374.77 \\
H$_2$ & S(2) &  12.28 &  1682 &   54.06 &   79.21 &   66.72 &  167.96 \\
H$_2$ & S(1) &  17.03 &  1015 &   33.46 &   51.38 &   46.35 &  126.41 \\
H$_2$ & S(0) &  28.22 &   510 &    7.23 &   10.28 &    8.28 &   18.60 \\
o-H$_2$O & 7$_{25}-6_{16}$ &  29.84 &  1126 &    0.55 &   ...   &   ...   & 
  ...   \\
o-H$_2$O & 6$_{34}-5_{05}$ &  30.90 &   934 &    0.79 &   ...   &   ...   & 
  ...   \\
o-H$_2$O & 7$_{34}-6_{25}$ &  34.55 &  1212 &    0.88 &   ...   &   ...   & 
  ...   \\
p-H$_2$O & 6$_{24}-5_{15}$ &  36.21 &   867 &    0.69 &   ...   &   ...   & 
  ...   \\
PII & $^3P_1- ^3P_0$ &  60.64 &   237 &   ...   &   ...   &    5.33 &   ...   \\
OI & $^3P_1- ^3P_2$ &  63.19 &   228 &   ...   &   ...   &  825.49 &  387.29 \\
o-H$_2$O & 3$_{21}-2_{12}$ &  75.38 &   305 &    9.35 &    3.41 &    3.04 & 
  ...   \\
o-H$_2$O & 4$_{23}-3_{12}$ &  78.74 &   432 &    3.93 &   ...   &   ...   & 
  ...   \\
OH & $^2\Pi_{1/2}-^2\Pi_{1/2}\,J=1/2^--3/2^+$ &  79.11 &   181 &    8.69 & 
   3.97 &    6.12 &   ...   \\
OH & $^2\Pi_{1/2}-^2\Pi_{3/2}\,J=1/2^+-3/2^-$ &  79.18 &   181 &    8.57 & 
   2.74 &    9.89 &   ...   \\
o-H$_2$O & 6$_{16}-5_{05}$ &  82.03 &   643 &    3.04 &   ...   &   ...   & 
  ...   \\
OH & $^2\Pi_{3/2}\,J=7/2^+-5/2^-$ &  84.42 &   291 &    4.82 &    4.67 &    6.11
 &   ...   \\
OH & $^2\Pi_{3/2}\,J=7/2^--5/2^+\, ^b$ &  84.60 &   290 &    6.96 &    3.54 & 
   4.82 &   ...   \\
CO & $J=30-29$ &  87.19 &  2565 &    3.88 &   ...   &   ...   &   ...   \\
OIII & $^3P_1- ^3P_0$ &  88.36 &   163 &   ...   &   ...   &   25.92 &   ...  
 \\
p-H$_2$O & 3$_{22}-2_{11}$ &  89.99 &   296 &    3.88 &   ...   &   ...   & 
  ...   \\
CO & $J=29-28$ &  90.16 &  2400 &    5.45 &    3.77 &   ...   &   ...   \\
CO & $J=28-27$ &  93.35 &  2240 &    3.31 &    1.71 &   ...   &   ...   \\
CO & $J=25-24$ & 104.44 &  1794 &    5.95 &    3.91 &    2.33 &   ...   \\
CO & $J=24-23$ & 108.76 &  1656 &    7.51 &    2.90 &    3.24 &   ...   \\
o-H$_2$O & $4_{14}-3_{03} \, ^c$ & 113.54 &   323 &   22.61 &    9.35 &    6.46
 &    2.53 \\
CO & $J=22-21$ & 118.58 &  1397 &    8.05 &    4.53 &    3.08 &   ...   \\
OH & $^2\Pi_{3/2}\,J=5/2^--3/2^+$ & 119.23 &   120 &    6.86 &    2.68 &    4.41
 &   -1.17 \\
OH & $^2\Pi_{3/2}\,J=5/2^+-3/2^-$ & 119.44 &   120 &    8.13 &    2.60 &    5.12
 &   -0.79$^d$ \\
NII & $^3P_2- ^3P_1$ & 121.89 &   188 &   ...   &   ...   &   29.81 &    8.34 \\
CO & $J=21-20$ & 124.19 &  1276 &   11.22 &    5.78 &    4.13 &    0.86 \\
p-H$_2$O & 4$_{04}-3_{13}$ & 125.35 &   319 &    4.00 &    1.33 &   ...   & 
  ...   \\
CO & $J=20-19$ & 130.37 &  1160 &   11.94 &   ...   &    4.62 &    0.45 \\
o-H$_2$O & 4$_{23}-4_{14}$ & 132.41 &   432 &    1.34 &   ...   &   ...   & 
  ...   \\
CO & $J=19-18$ & 137.20 &  1050 &   15.29 &    6.86 &    5.41 &    0.77 \\
p-H$_2$O & 3$_{13}-2_{02}$ & 138.53 &   204 &   11.75 &    3.32 &    2.59 & 
   1.19 \\
CO & $J=18-17$ & 144.78 &   945 &   17.17 &    6.85 &    8.85 &    1.46 \\
OI & $^3P_0- ^3P_1$ & 145.53 &   188 &    9.65 &   12.90 &   98.76 &   50.52 \\
CO & $J=17-16$ & 153.27 &   846 &   19.53 &    8.55 &    9.39 &    2.40 \\
p-H$_2$O & 3$_{22}-3_{13}$ & 156.19 &   296 &    1.60 &   ...   &   ...   & 
  ...   \\
CII & $^2P_{3/2} - ^2P_{1/2} \,^e$ & 157.74 &    92 &  -24.93 &   62.79 & 
 202.98 &  166.46 \\
CO & $J=16-15$ & 162.81 &   752 &   29.04 &   ...   &   12.57 &    2.89 \\
OH & $^2\Pi_{1/2}\,J=3/2^+-1/2^-$ & 163.12 &   270 &    2.45 &   ...   &    2.25
 &    0.81 \\
OH & $^2\Pi_{1/2}\,J=3/2^--1/2^+$ & 163.40 &   269 &    1.35 &    0.35 &    1.30
 &    0.95 \\
CO & $J=15-14$ & 173.63 &   663 &   27.69 &   12.16 &   13.14 &    3.78 \\
o-H$_2$O & 3$_{03}-2_{12}$ & 174.62 &   196 &   23.92 &    6.46 &    5.64 & 
   2.66 \\
o-H$_2$O & 2$_{12}-1_{01}$ & 179.53 &   114 &   27.20 &    8.55 &    8.66 & 
   2.49 \\
o-H$_2$O & 2$_{21}-2_{12}$ & 180.49 &   194 &    8.26 &    1.81 &    0.82 & 
  ...   \\
CO & $J=14-13$ & 186.00 &   580 &   29.34 &   11.67 &   12.81 &    3.95 \\

\enddata

\tablenotetext{a}{For strong lines, uncertainties in the flux calibration of the PACS instrument lead to r.m.s. errors of $11-12 \%$ and peak-to-peak errors of $\pm 30\%$.  For the observations of W28 and 3C391, the 1 sigma statistical errors on the line fluxes are approximately $4 \times 10^{-17}$, $1.0 \times 10^{-17}$, and $1.5 \times 10^{-17}$,~W~m$^{-2}$ respectively for the 70 -- 95, 100 -- 140,   and 140 -- 190~$\mu$m spectral regions.  For the observations of NGC 2071 (both positions), the 1 sigma statistical errors are approximately $4 \times 10^{-17}$, $1.5 \times 10^{-17}$, and $1.5 \times 10^{-17}$~W~m$^{-2}$ respectively for the 70 -- 95, 100 -- 140,  and 140 -- 190~$\mu$m spectral regions.  Multiplying these values by 3 yields the 3 sigma upper limit on any undetected line within these spectral windows.}

\tablenotetext{b}{Blend with CO $J=31-30$}

\tablenotetext{c}{Blend with CO $J=23-22$}

\tablenotetext{d}{Absorption line}

\tablenotetext{e}{Fluxes strongly affected by emission extending to the reference beam.  (In NGC 2071 NE, the CII flux is larger at the reference positions than at the source position, leading to a negative entry.) }

\vskip 0.2 true in
\end{deluxetable}

\begin{deluxetable}{lcccc}
\tablecaption{Best-fit parameters}
\tablehead{  				&  NGC2071 NE     & NGC2071 SW      & 3C391   & W28      \\
& \phantom{0000000000000} &\phantom{0000000000000} &\phantom{0000000000000} &\phantom{0000000000000} \\ }  
\startdata

log$_{10}\, [n({\rm H}_2)/{\rm cm}^{-3}]$ 	& \phantom{--}5.0 			& \phantom{--}4.6		& \phantom{--}4.4		& 3.6 \\
Power-law index, $b$		& \phantom{--}2.9			& \phantom{--}2.9 		& \phantom{--}2.7     	& 3.2 \\
OPR$_0$	$^b$					& \phantom{--}0.9   		& \phantom{--}0.9		& \phantom{--}0.9		& 1.3 \\	
log$_{10}\, [\tau n({\rm H}) \rm \, in \, \, cm^{-3}\,s]$ $^c$	& \phantom{--}11.0  		& \phantom{--}11.1		& \phantom{--}12.0		& 12.0 \\
log$_{10}\, (T_{\rm w}/K)$	    & \phantom{--}2.5 			& \phantom{--}2.3		& \phantom{--}2.4       & 3.1 \\
log$_{10}\, [n({\rm H_2O})/n({\rm CO})]$& --0.6 		& --1.0		& --0.9		& 0.1 \\
\re{log$_{10}\, (T_{\rm w}/K)^a$} & 3.0 & 3.1 & 3.2 & 3.1 \\
\enddata
\tablenotetext{a}{\re{Best-fit value obtained with $n({\rm H_2O})/n({\rm CO})=1.2$ assumed as a prior}}
\tablenotetext{b}{\bl{Initial H$_2$ ortho-to-para ratio}}
\tablenotetext{c}{\bl{Product of the H atom density and the time period for ortho-para conversion}}
\end{deluxetable}

\vfill\eject

\begin{figure} 
\includegraphics[width=14 cm, angle=0]{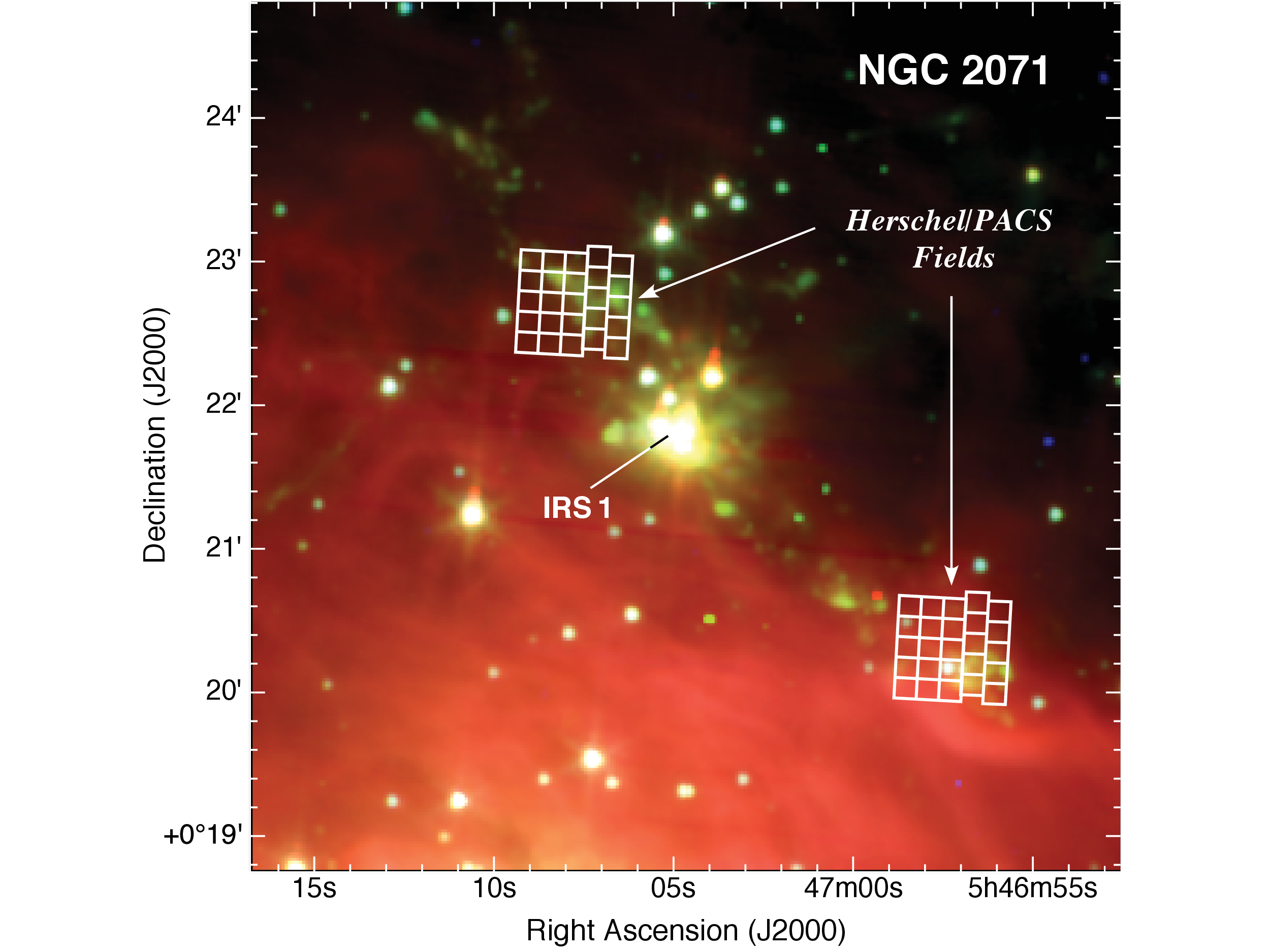}
\caption{Finder chart for NGC 2071, showing a map of a $0.1^0 \times 0.1^0$ region centered on the outflow source NGC 2071 observed by the {\it Spitzer}/IRAC instrument. The emission in IRAC Bands 1, 2, and 4 is shown (on a log-scale) in blue, green, and red, respectively.  White squares indicate the positions of the {\it Herschel}/PACS spaxels in the observations reported here towards the northeast and southwest outflow lobes.}
\end{figure}

\begin{figure} 
\includegraphics[width=14 cm, angle=0]{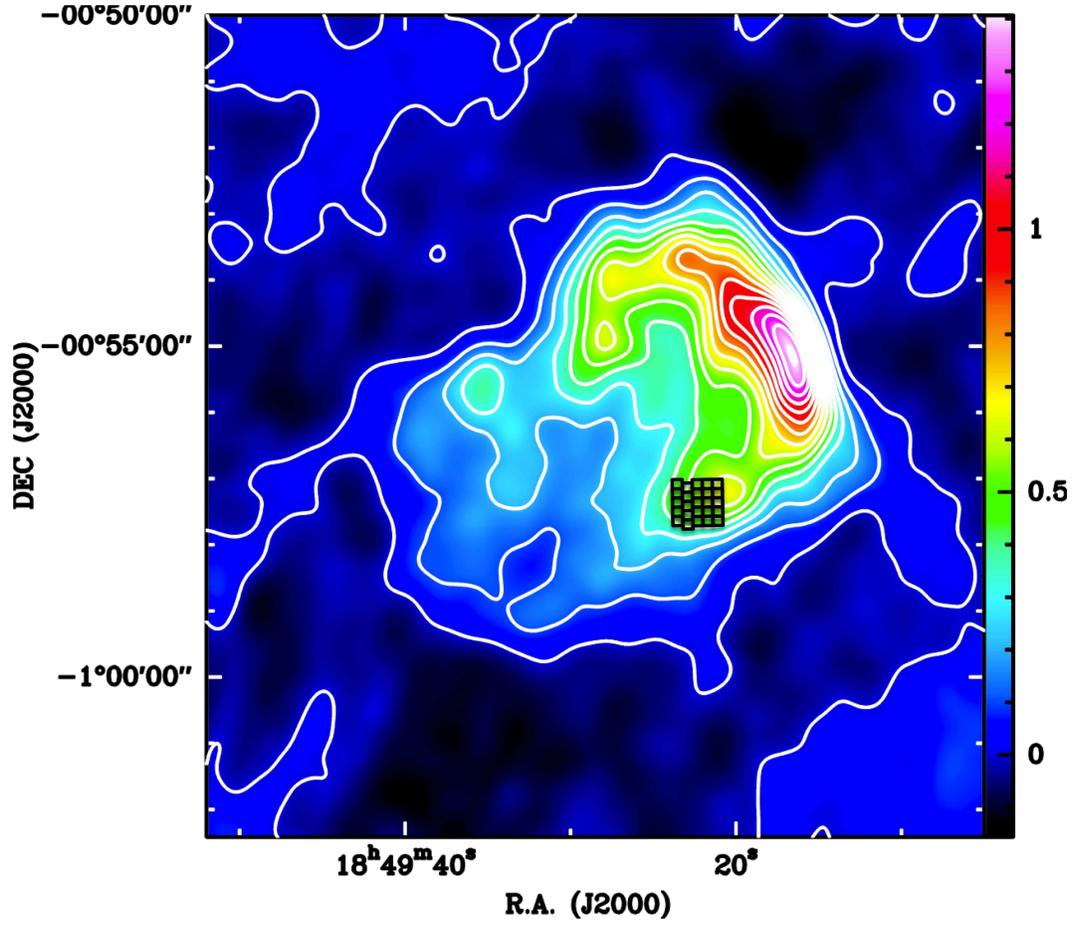}
\caption{Finder chart for 3C391, showing a map of the 330 MHz radio continuum emission (Moffett et al.\ 1994, in units of Jy/beam).    Black squares indicate the positions of the {\it Herschel}/PACS spaxels in the observations reported here.}
\end{figure}

\begin{figure} 
\includegraphics[width=16 cm, angle=0]{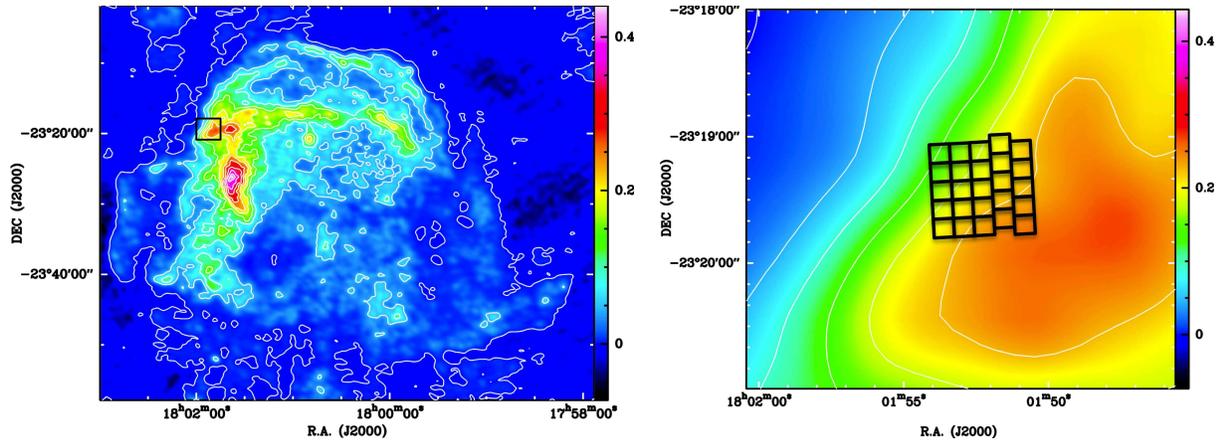}
\caption{Finder chart for W28, showing a map of the 327 MHz radio continuum emission (Claussen et al.\ 1997, in units of Jy/beam).  Black squares indicate the positions of the {\it Herschel}/PACS spaxels in the observations reported here. }
\end{figure}

\begin{figure} 
\includegraphics[width=14 cm, angle=0]{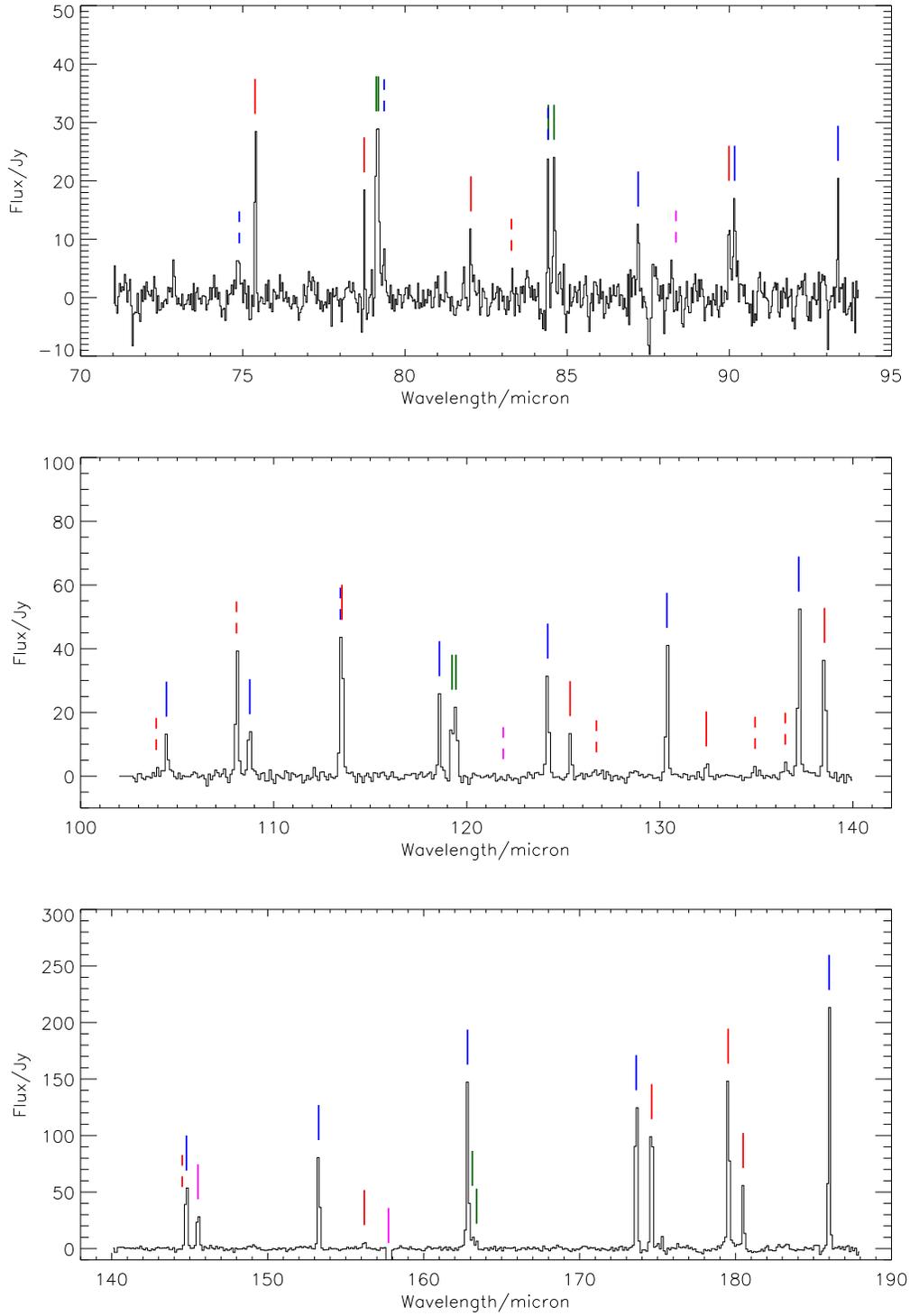}
\caption{PACS spectra obtained toward NGC 2071 NE.  Line positions are shown for rotational lines of H$_2$O (red), CO (blue), OH (green) and for fine-structure transitions of atoms and atomic ions (magenta).}
\end{figure}

\begin{figure} 
\includegraphics[width=14 cm, angle=0]{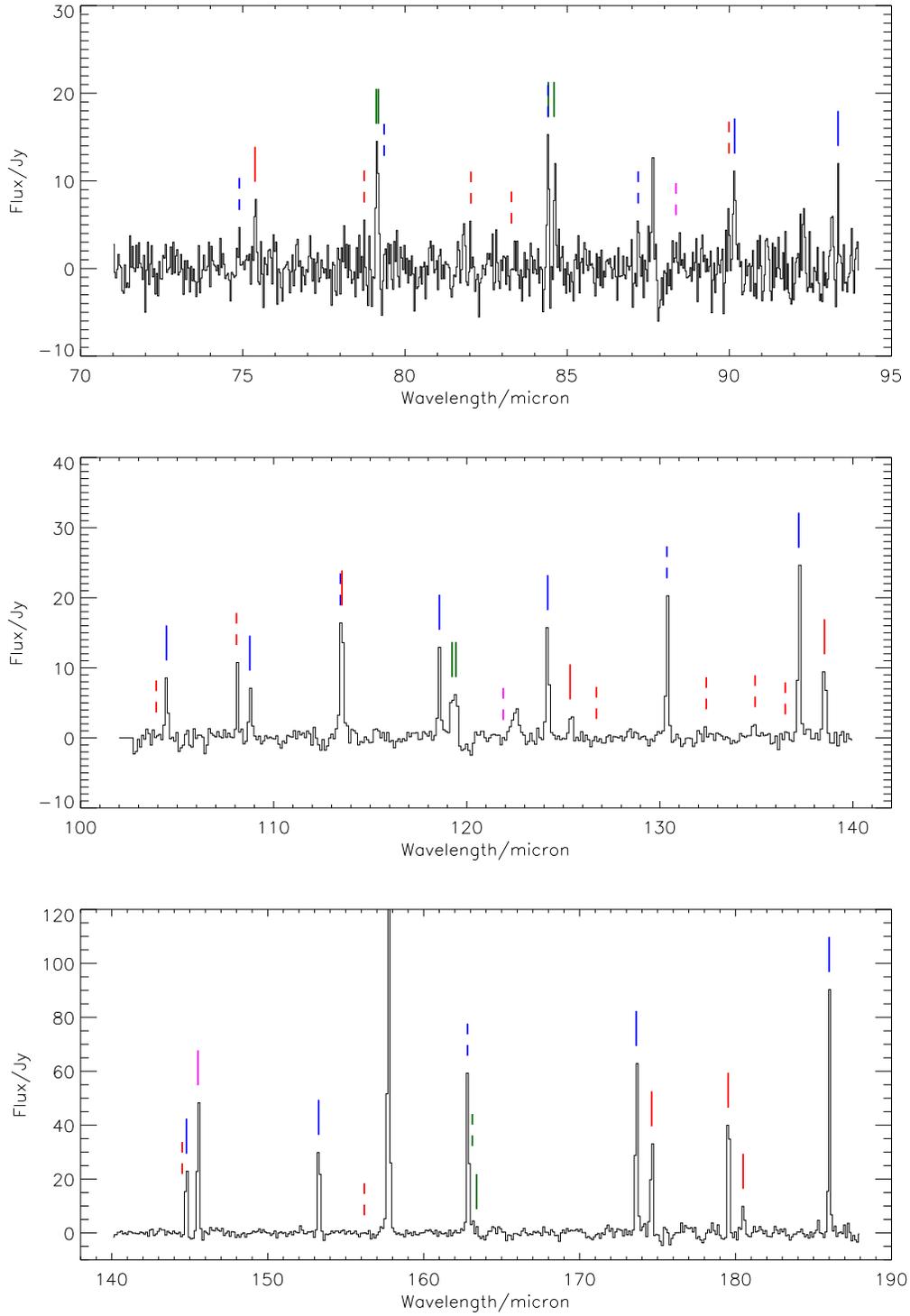}
\caption{PACS spectra obtained toward NGC 2071 SW.  Line positions are shown for rotational lines of H$_2$O (red), CO (blue), OH (green) and for fine-structure transitions of atoms and atomic ions (magenta).  The broad feature at 123~$\mu$m is an artifact.}
\end{figure}

\begin{figure} 
\includegraphics[width=14 cm, angle=0]{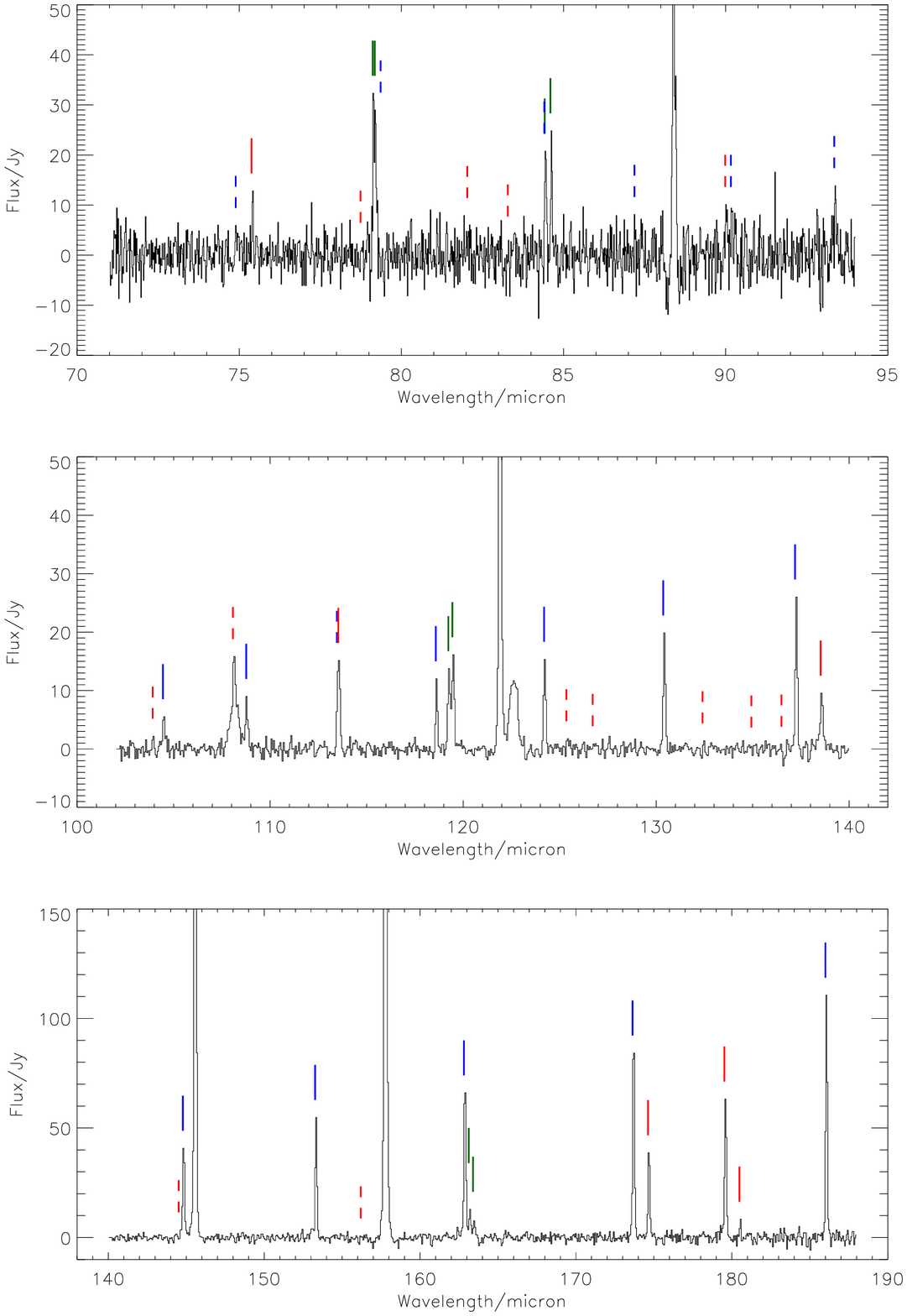}
\caption{PACS spectra obtained toward 3C391.  Line positions are shown for rotational lines of H$_2$O (red), CO (blue), OH (green) and for fine-structure transitions of atoms and atomic ions (magenta).  The broad features at 108 and 123~$\mu$m are artifacts.}
\end{figure}

\begin{figure} 
\includegraphics[width=14 cm, angle=0]{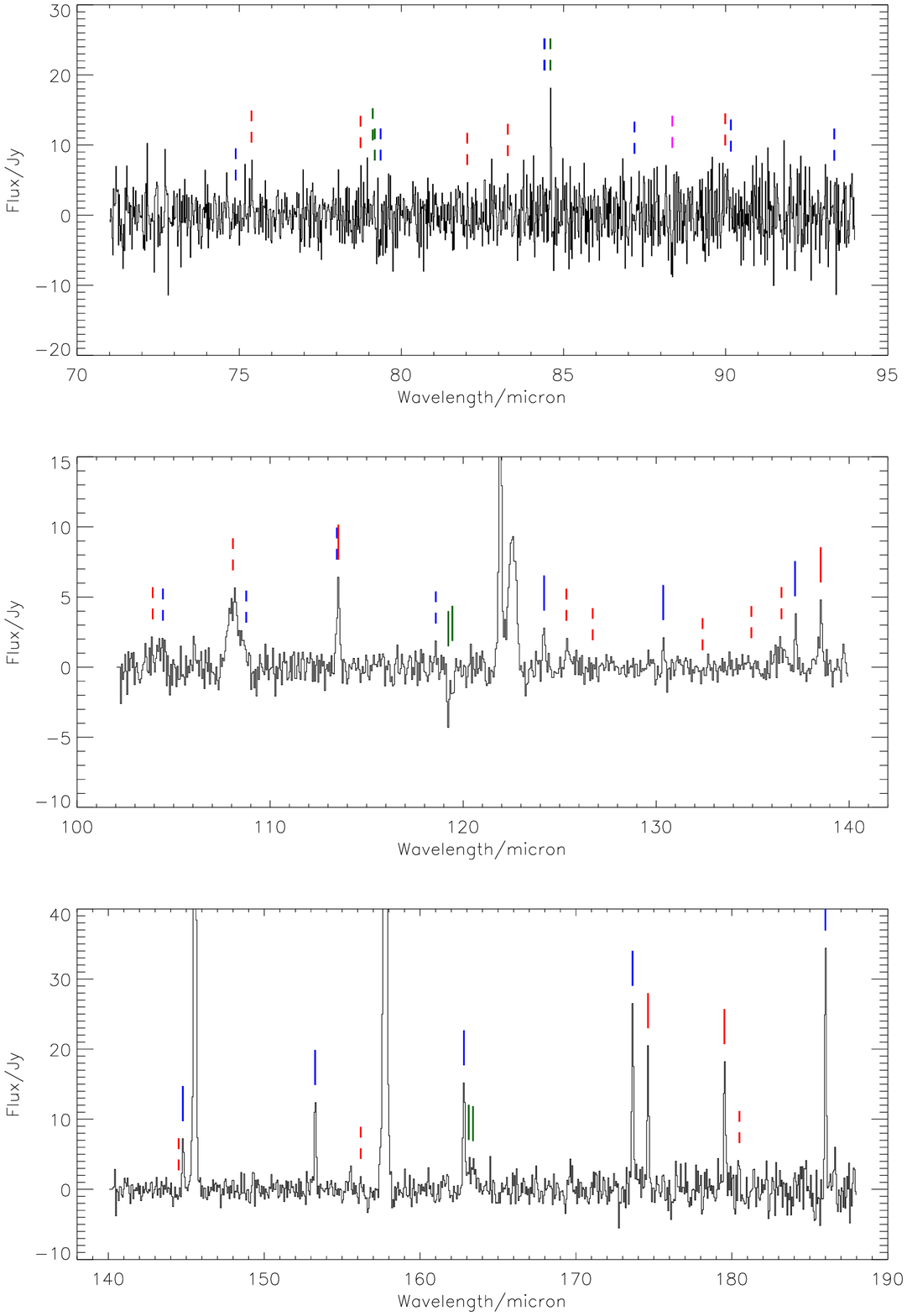}
\caption{PACS spectra obtained toward W28.  Line positions are shown for rotational lines of H$_2$O (red), CO (blue), OH (green) and for fine-structure transitions of atoms and atomic ions (magenta).  The broad features at 108 and 123~$\mu$m are artifacts.}
\end{figure}

\begin{figure} 
\includegraphics[width=13 cm, angle=0]{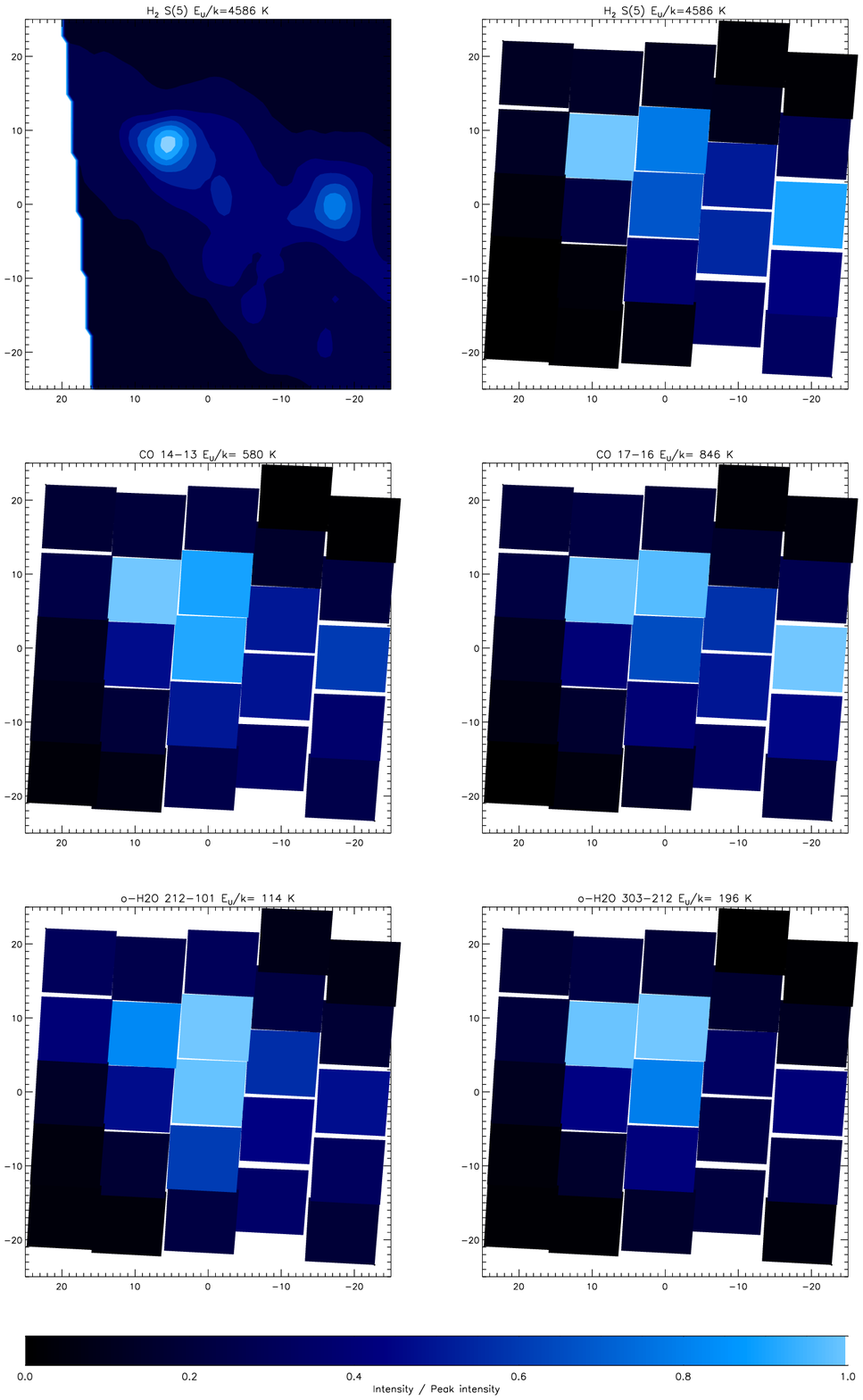}
\caption{Spatial distribution of line emissions in NGC 2071 NE}
\end{figure}

\begin{figure} 
\includegraphics[width=13 cm, angle=0]{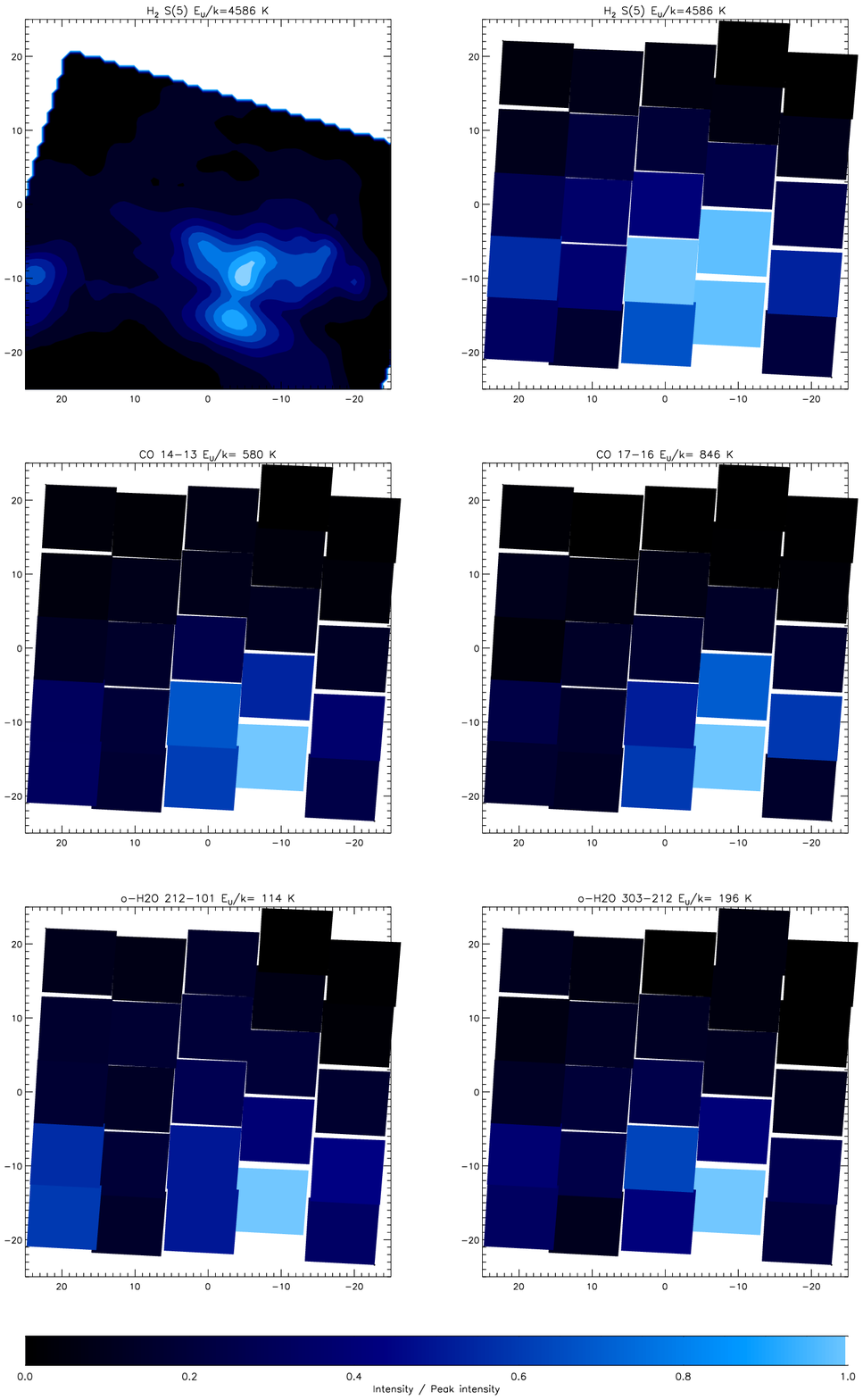}
\caption{Spatial distribution of line emissions in NGC 2071 SW}
\end{figure}

\begin{figure} 
\includegraphics[width=13 cm, angle=0]{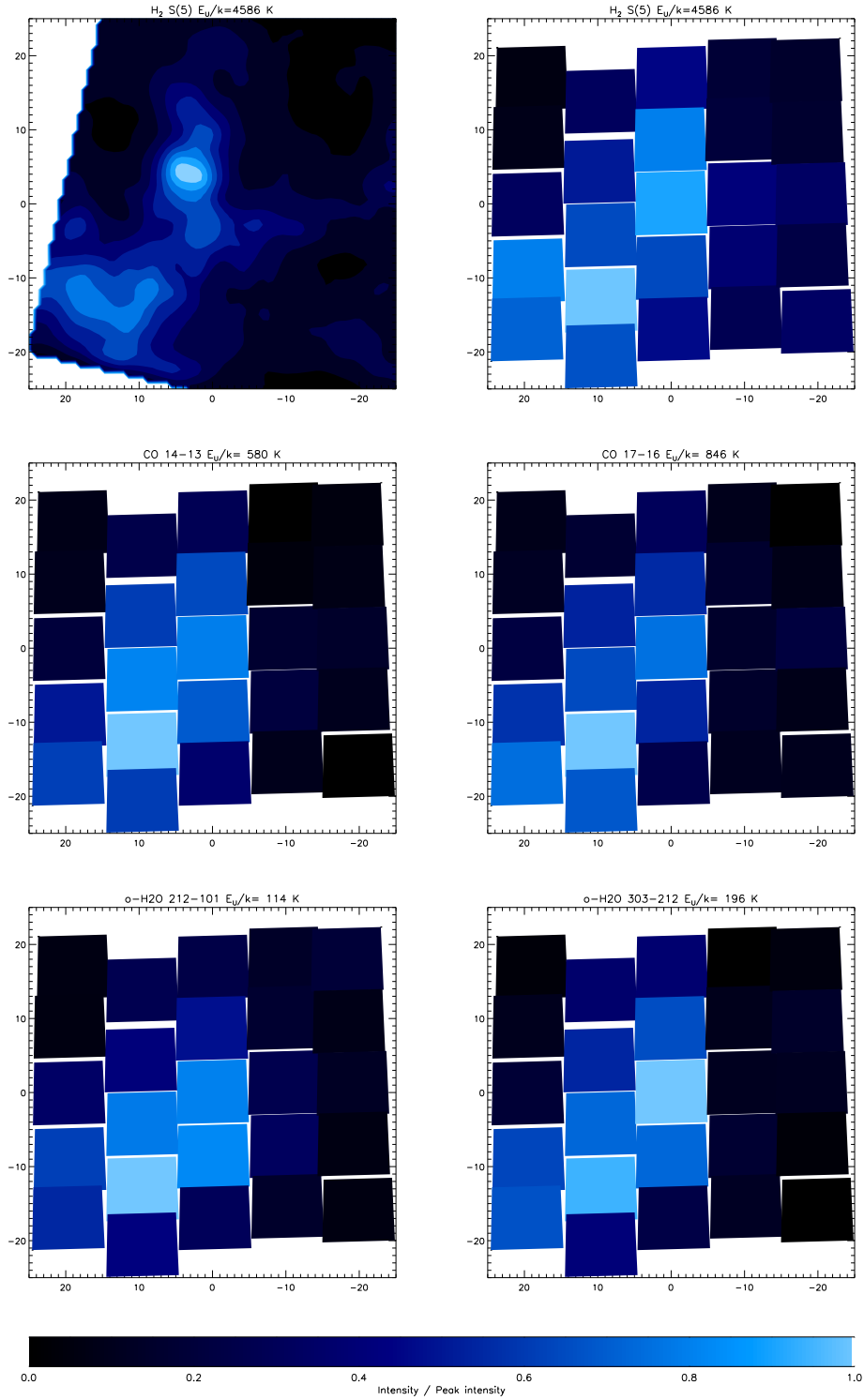}
\caption{Spatial distribution of line emissions in 3C391}
\end{figure}

\begin{figure} 
\includegraphics[width=13 cm, angle=0]{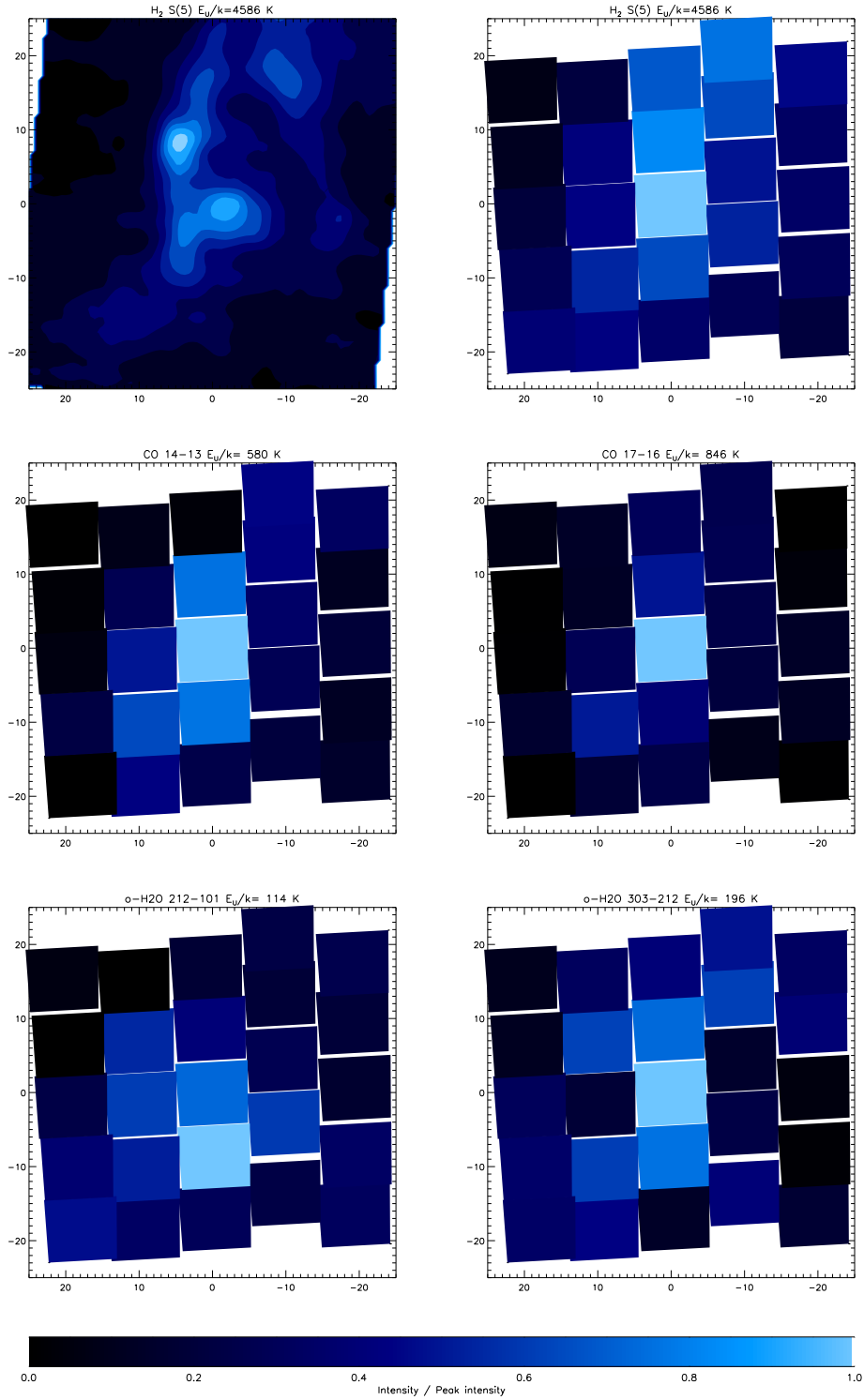}
\caption{Spatial distribution of line emissions in W28}
\end{figure}

\begin{figure} 
\includegraphics[width=7 cm, angle=0]{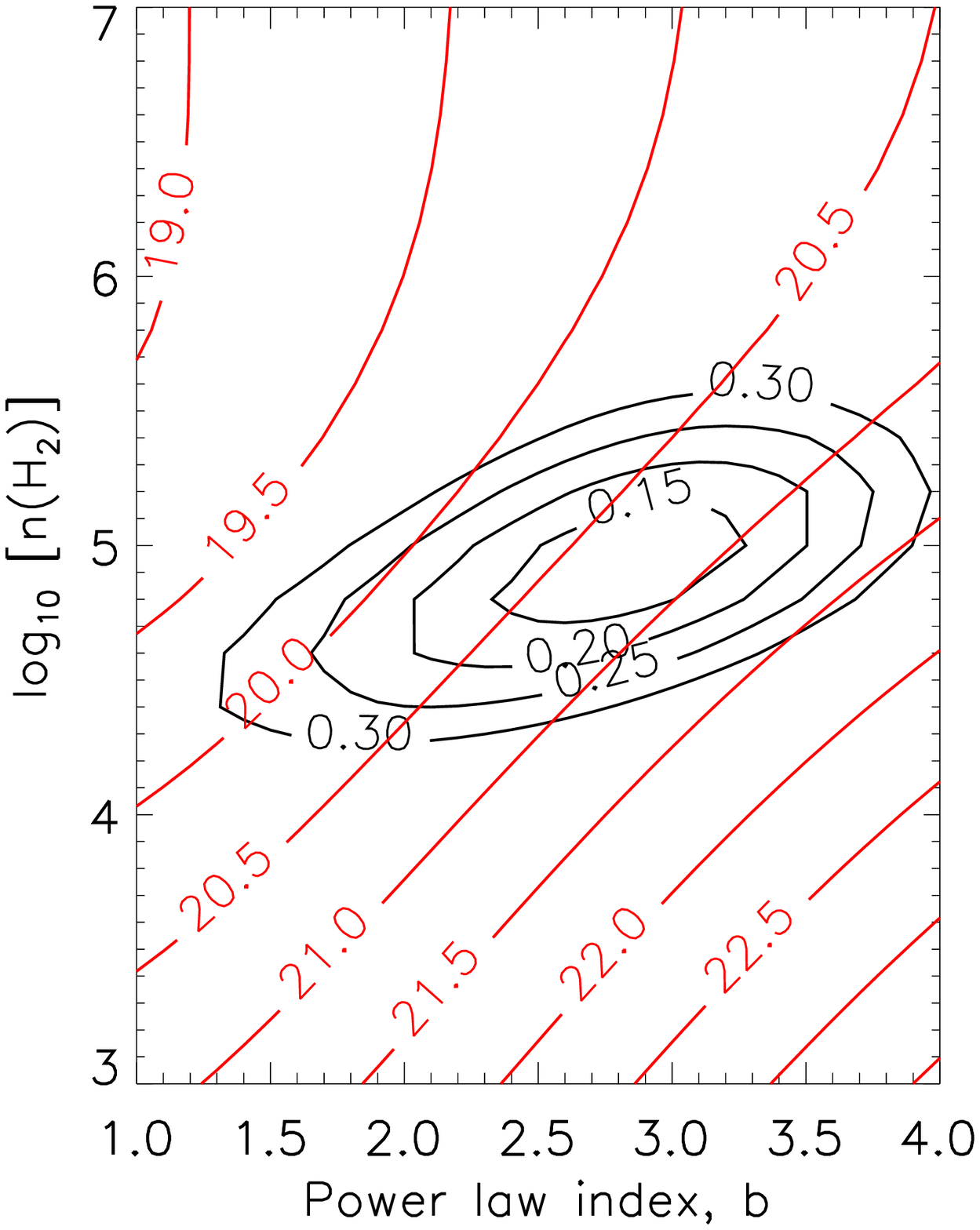}%
\includegraphics[width=7 cm, angle=0]{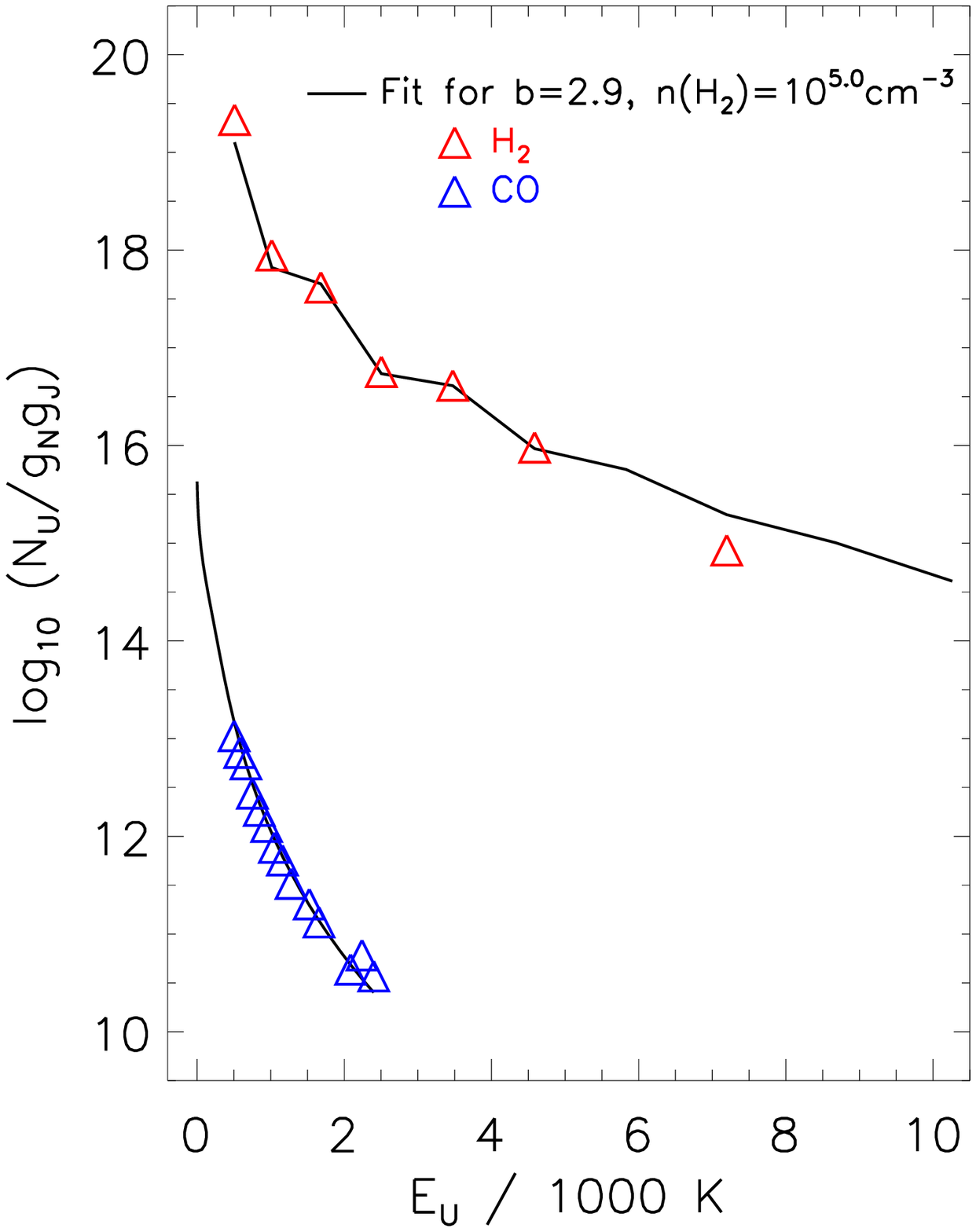}%

\includegraphics[width=7 cm, angle=0]{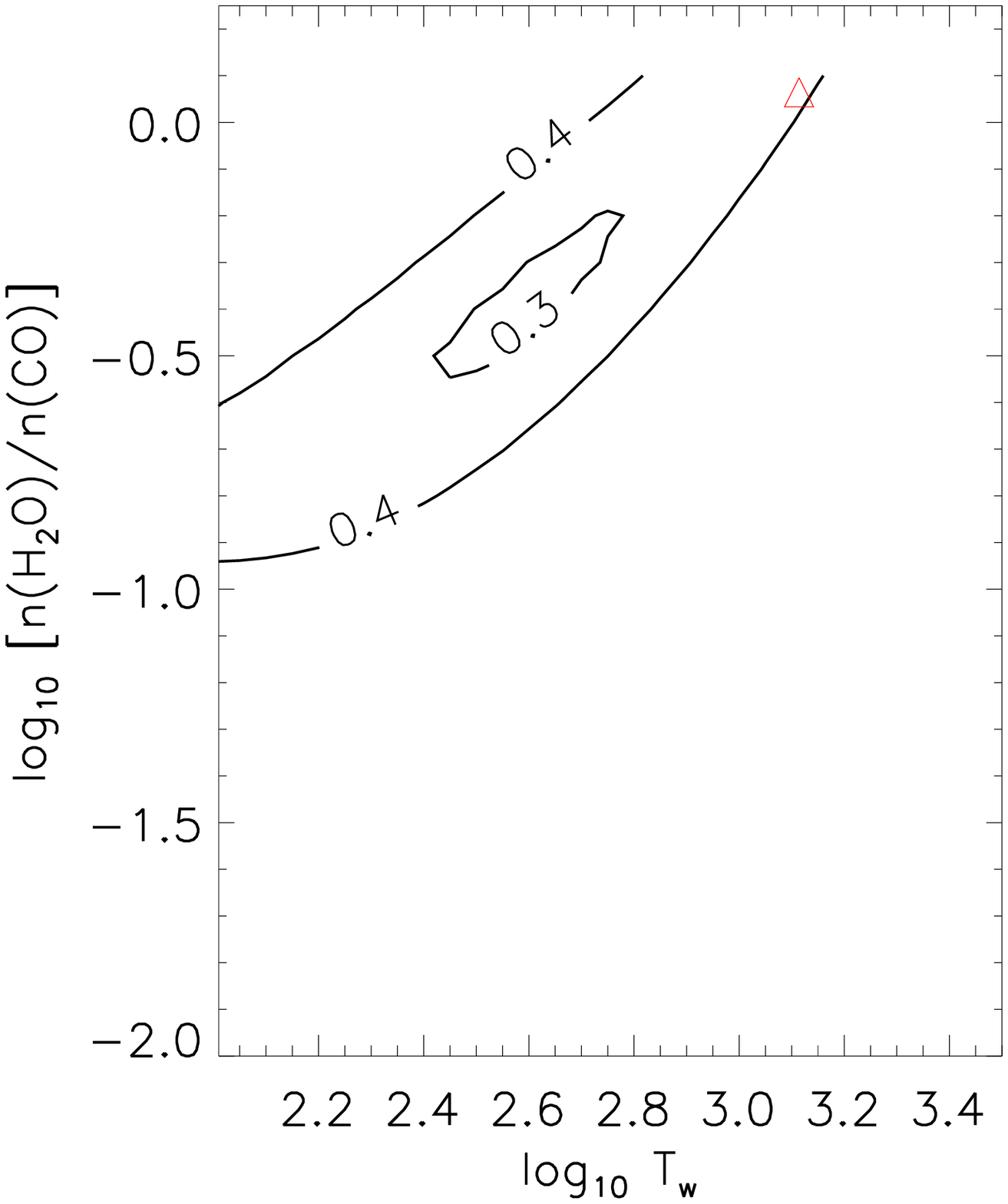}%
\includegraphics[width=7 cm, angle=0]{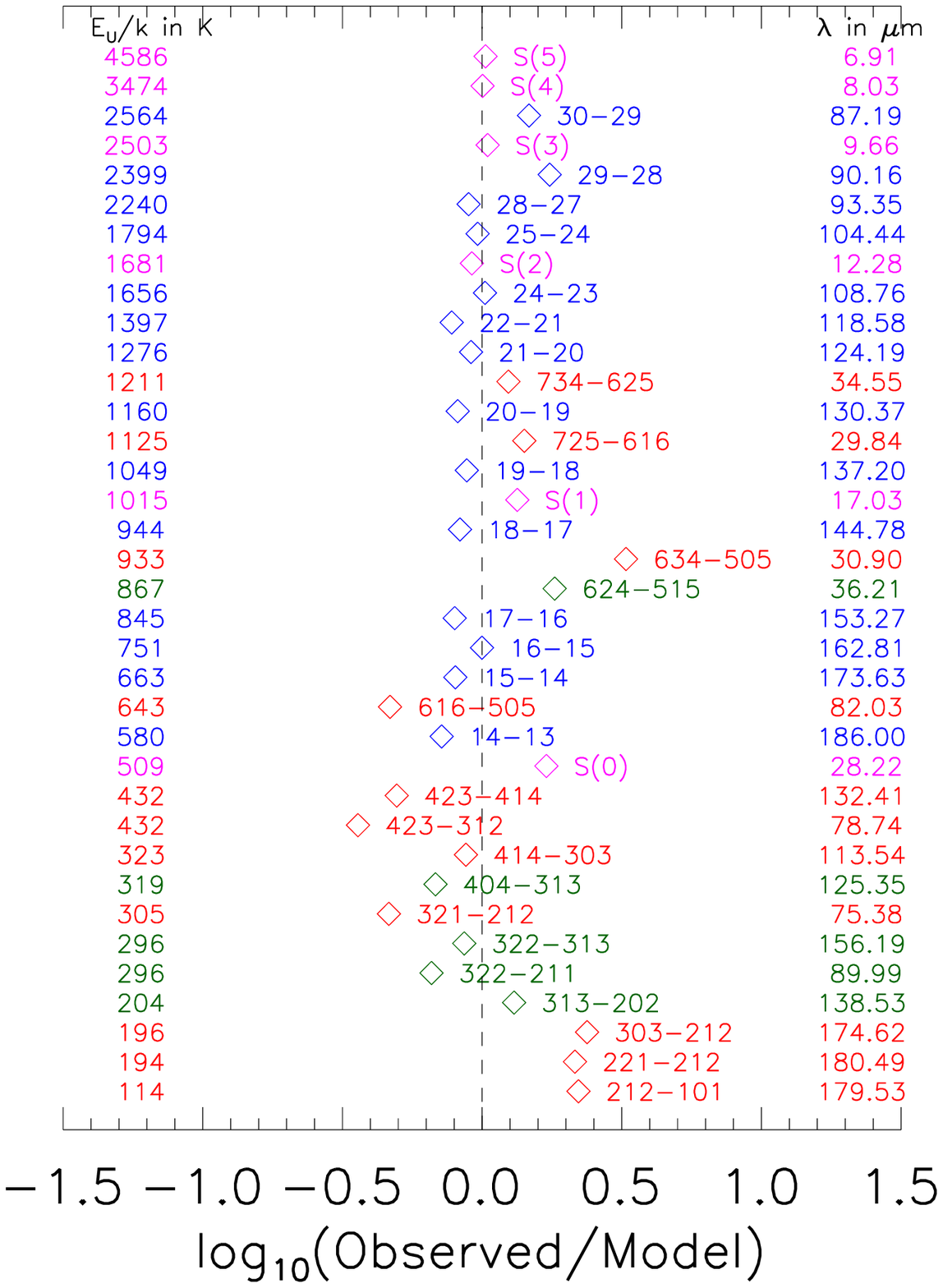}%                                        
\caption{Fit to the line fluxes observed from NGC 2071 NE (see text \bl{ for full explanation).  Top left:  goodness-of-fit to H$_2$ and CO line fluxes, as a function of H$_2$ density and powerlaw index, b, (black contours), and corresponding beam-averaged ${\rm log}_{10}(N({\rm H}_2)/\rm cm^{-2})$ (red contours).  Top right: Rotational diagrams for CO (blue symbols) and H$_2$ (red symbols), with fits (black curves).  Bottom left: goodness-of-fit for the H$_2$O line fluxes, as a function of the H$_2$O/CO ratio and the threshold temperature, $T_{\rm w}$.  Bottom right: Ratio of observed to fitted line flux for all transitions in best fit.}
} 
\end{figure}

\begin{figure} 
\includegraphics[width=8 cm, angle=0]{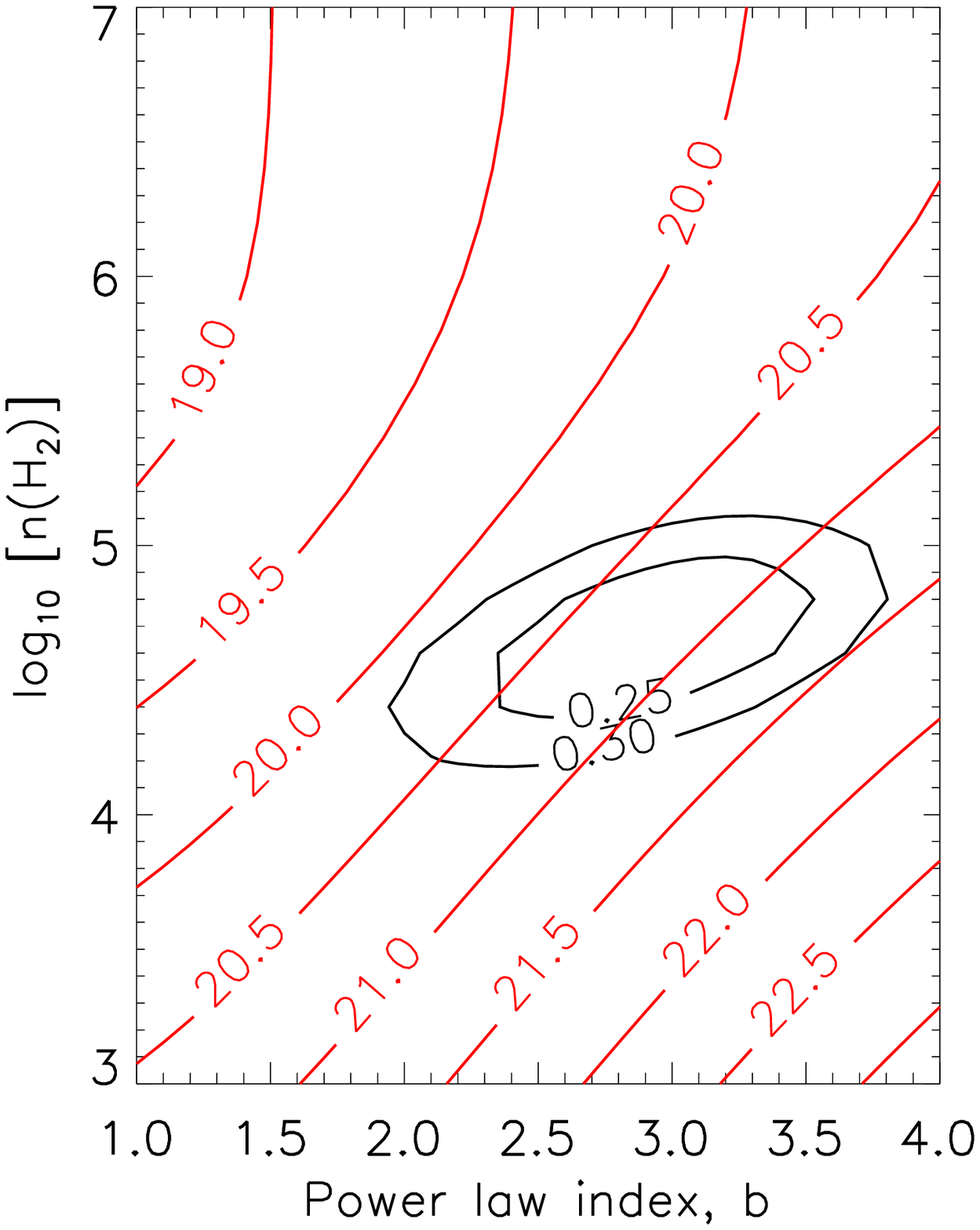}%
\includegraphics[width=8 cm, angle=0]{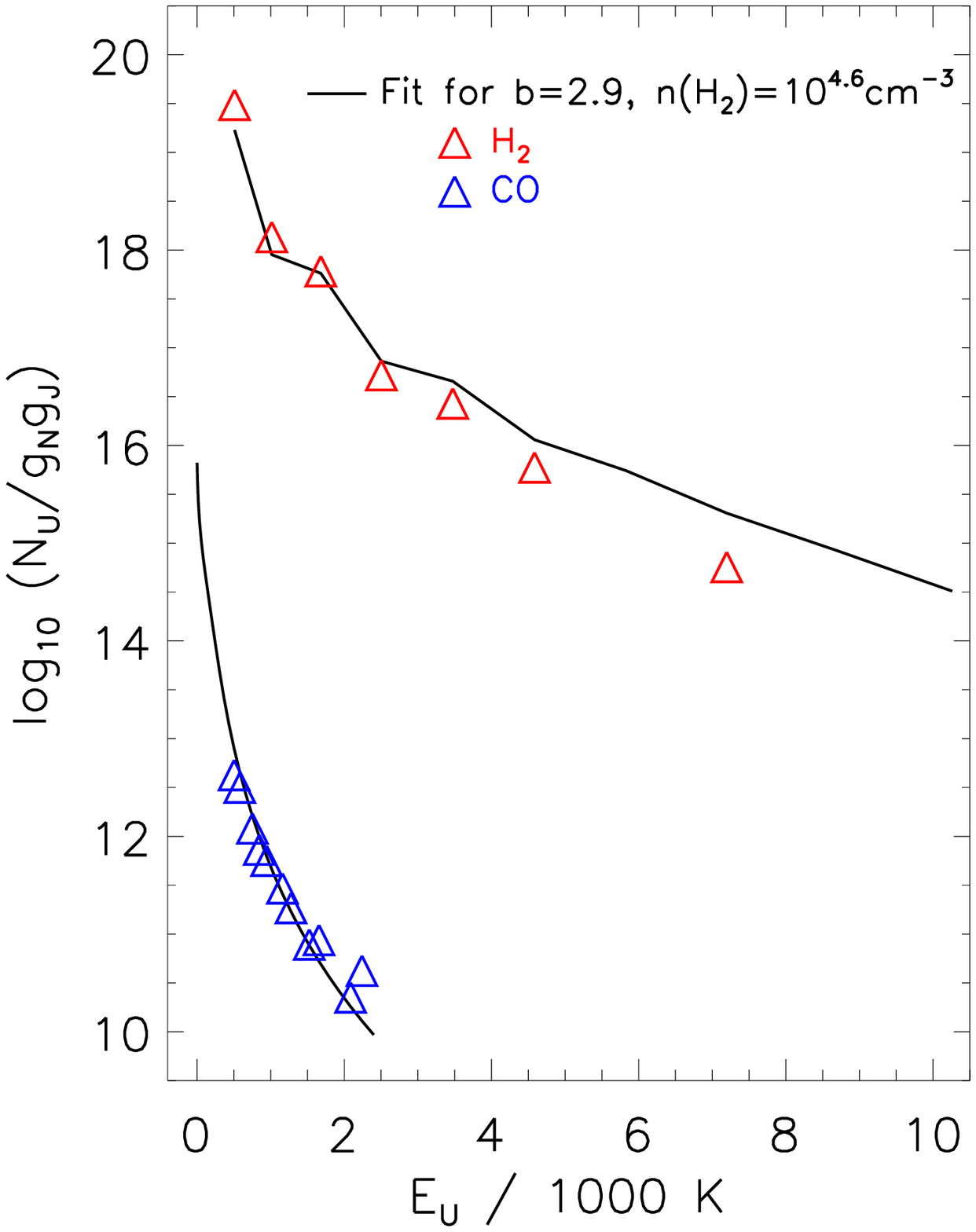}%

\includegraphics[width=8 cm, angle=0]{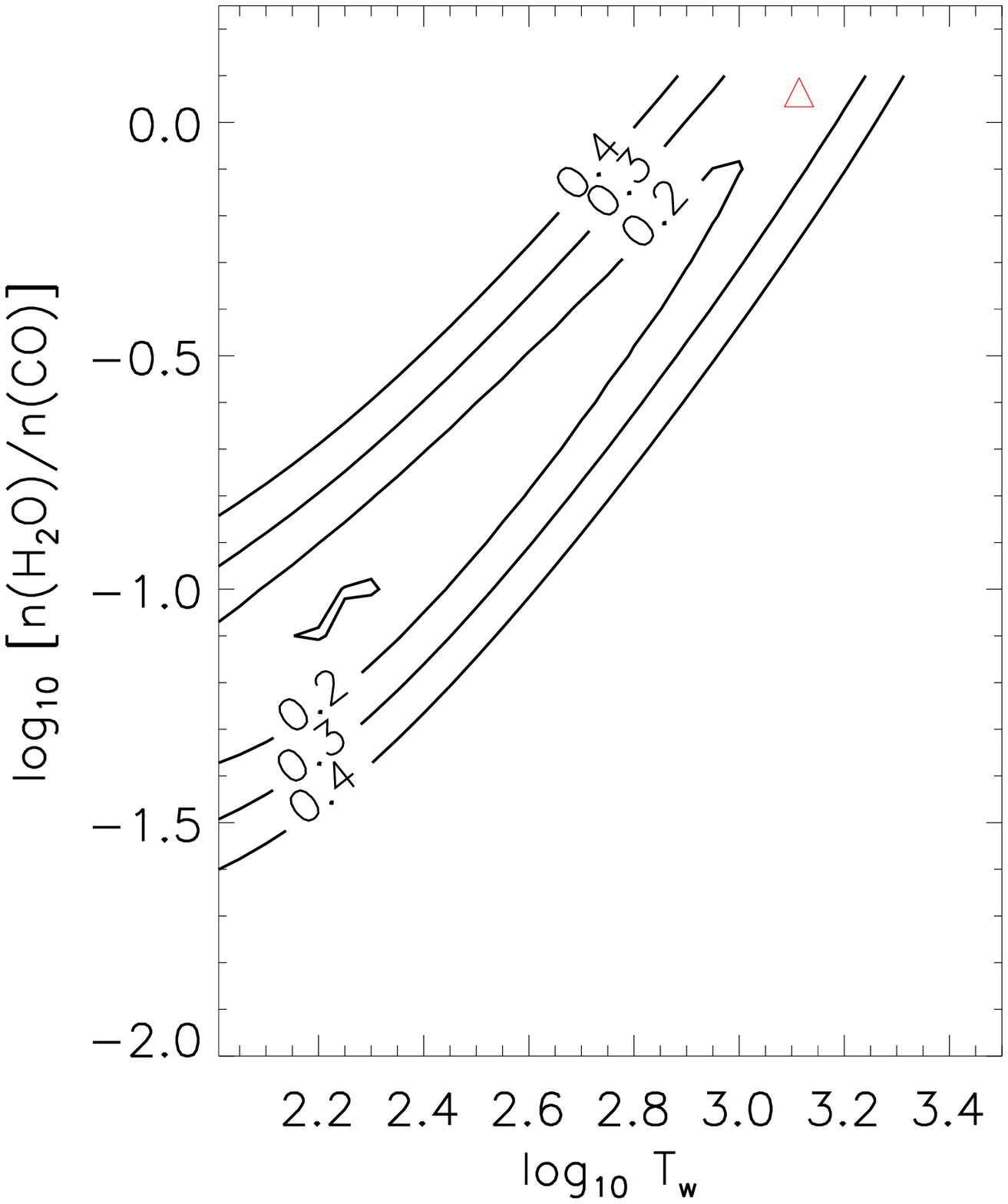}%
\includegraphics[width=8 cm, angle=0]{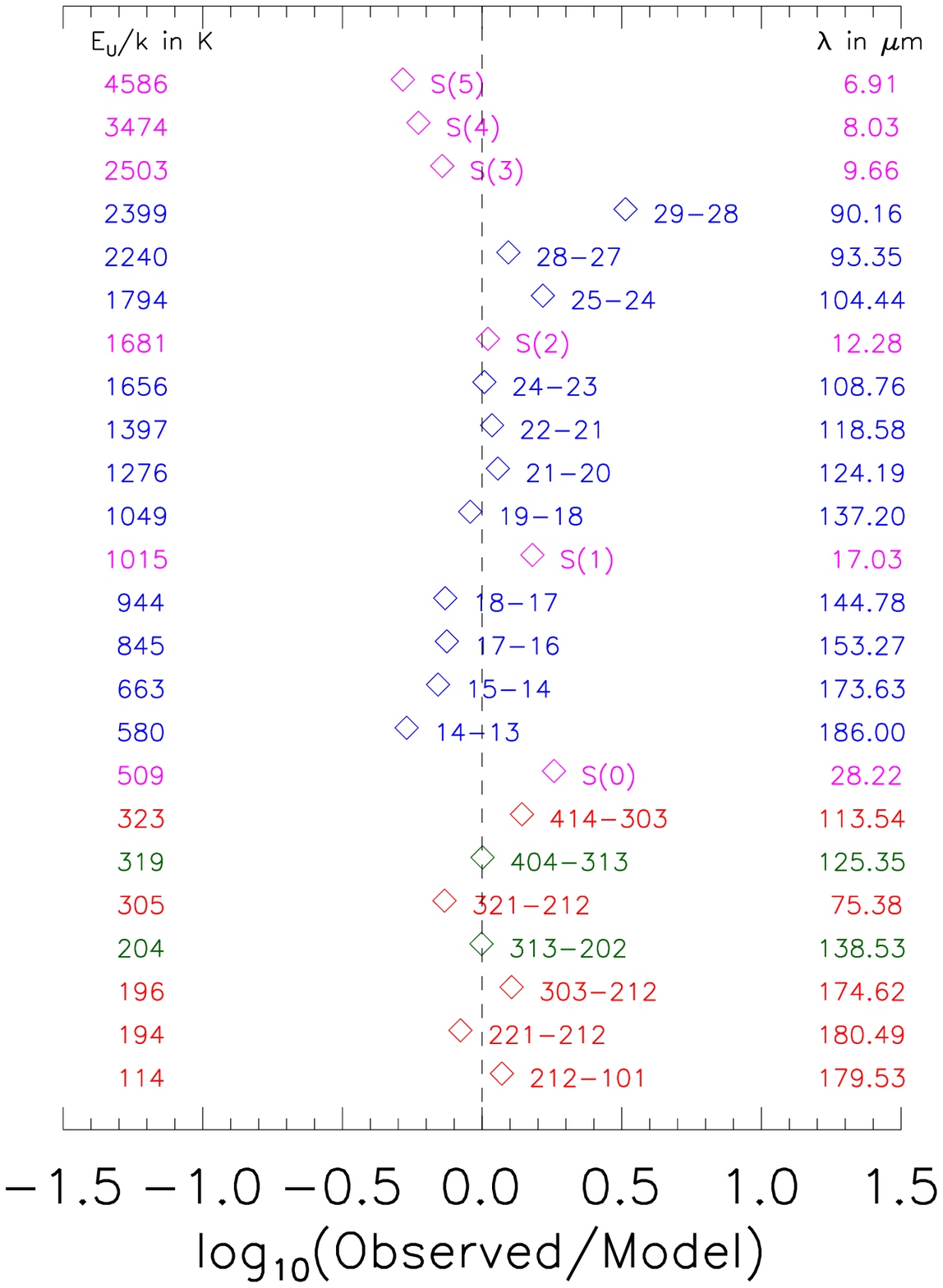}%                                        
\caption{Fit to the line fluxes observed from NGC 2071 SW (see text and Figure 12 caption for explanation)} 
\end{figure}

\begin{figure} 
\includegraphics[width=8 cm, angle=0]{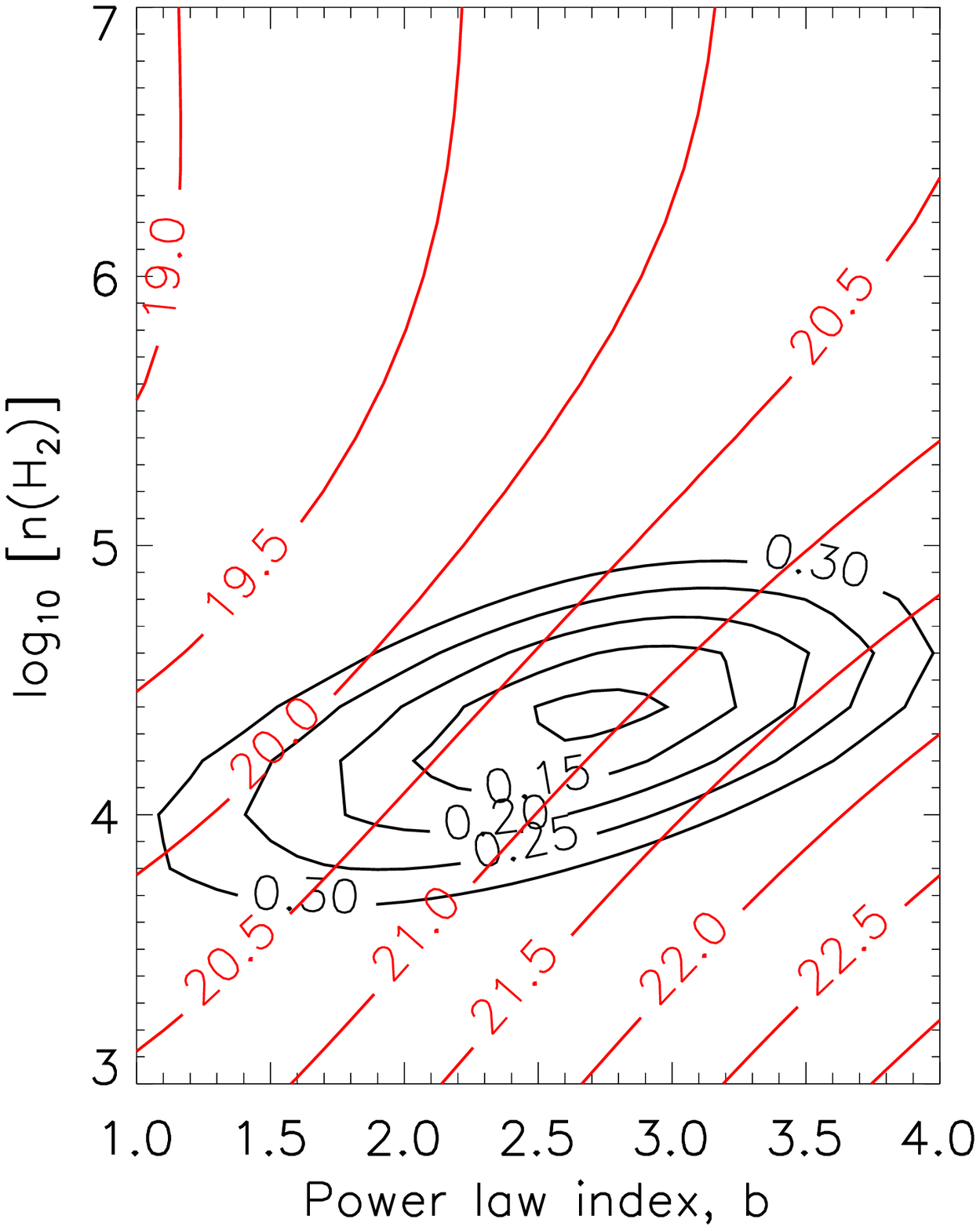}%
\includegraphics[width=8 cm, angle=0]{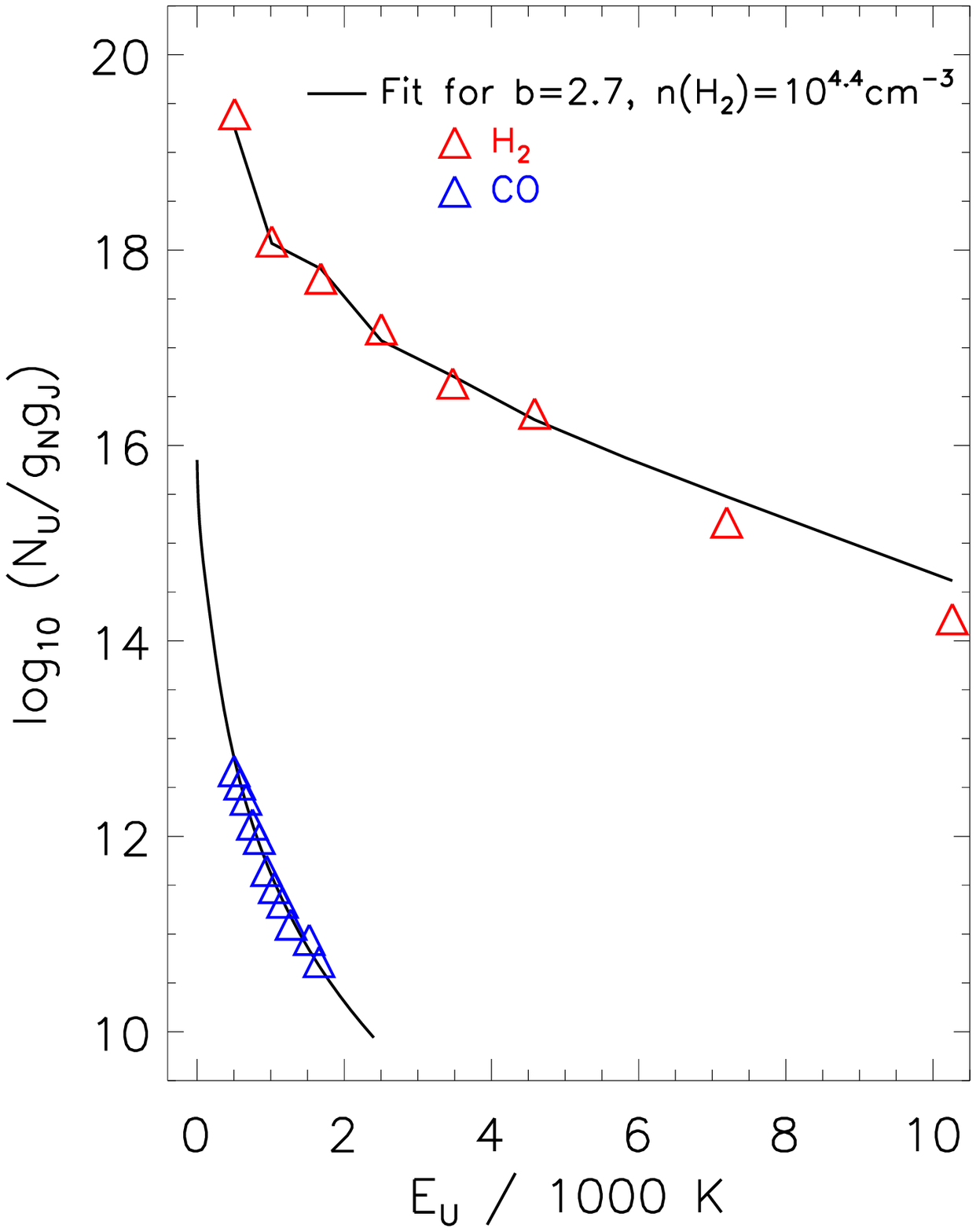}%

\includegraphics[width=8 cm, angle=0]{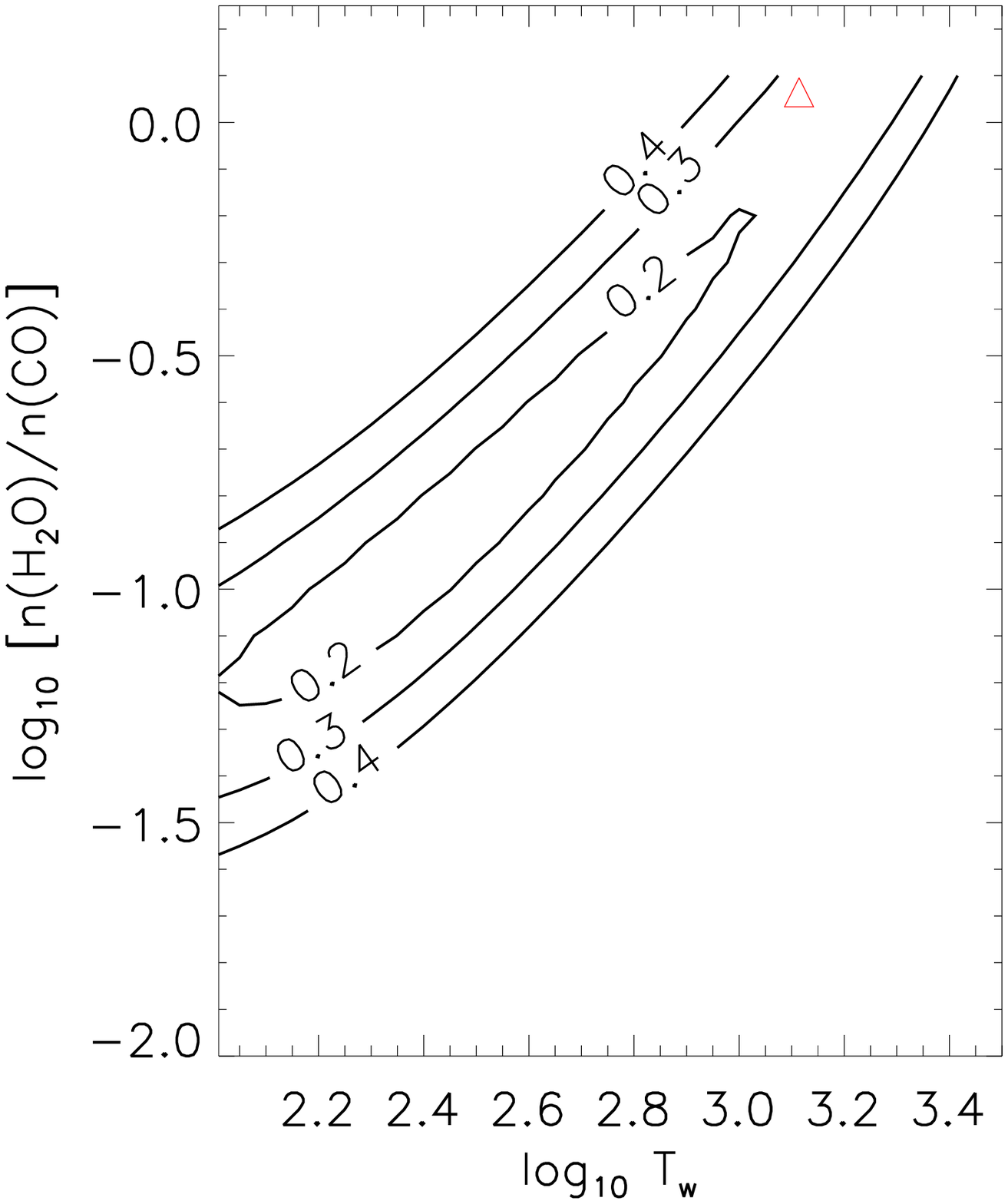}%
\includegraphics[width=8 cm, angle=0]{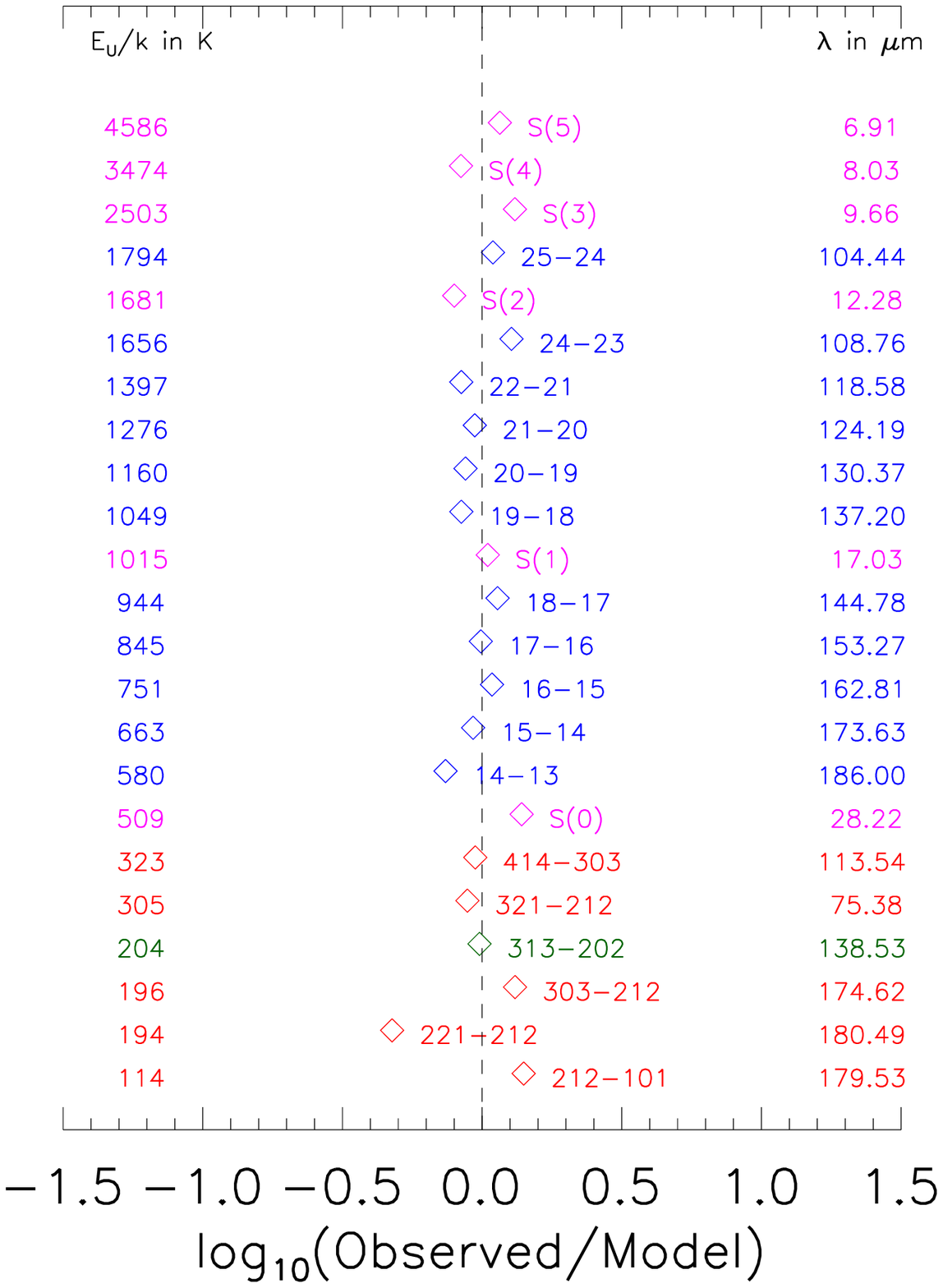}%                                        
\caption{Fit to the line fluxes observed from 3C391 (see text and Figure 12 caption for explanation)} 
\end{figure}

\begin{figure} 
\includegraphics[width=8 cm, angle=0]{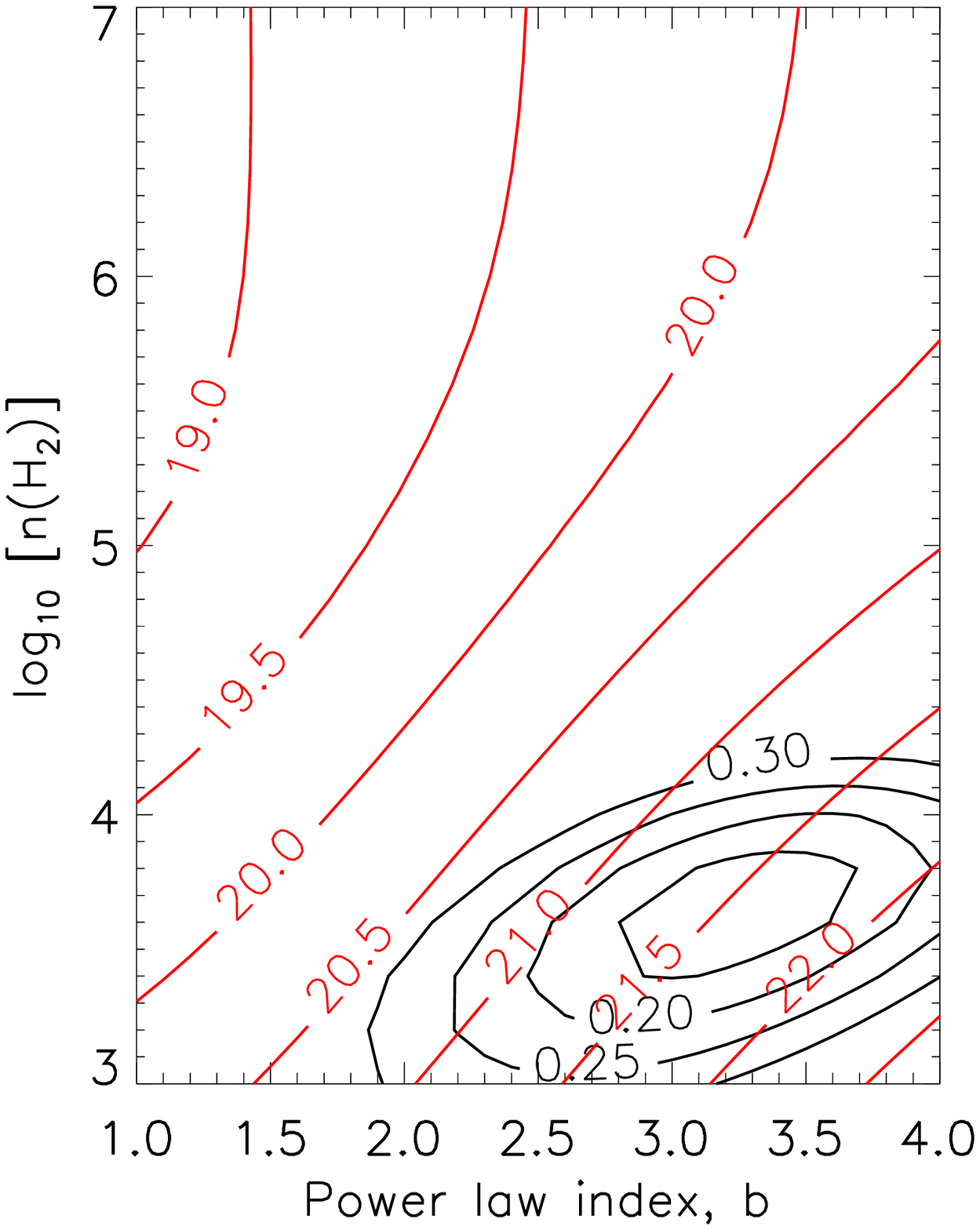}%
\includegraphics[width=8 cm, angle=0]{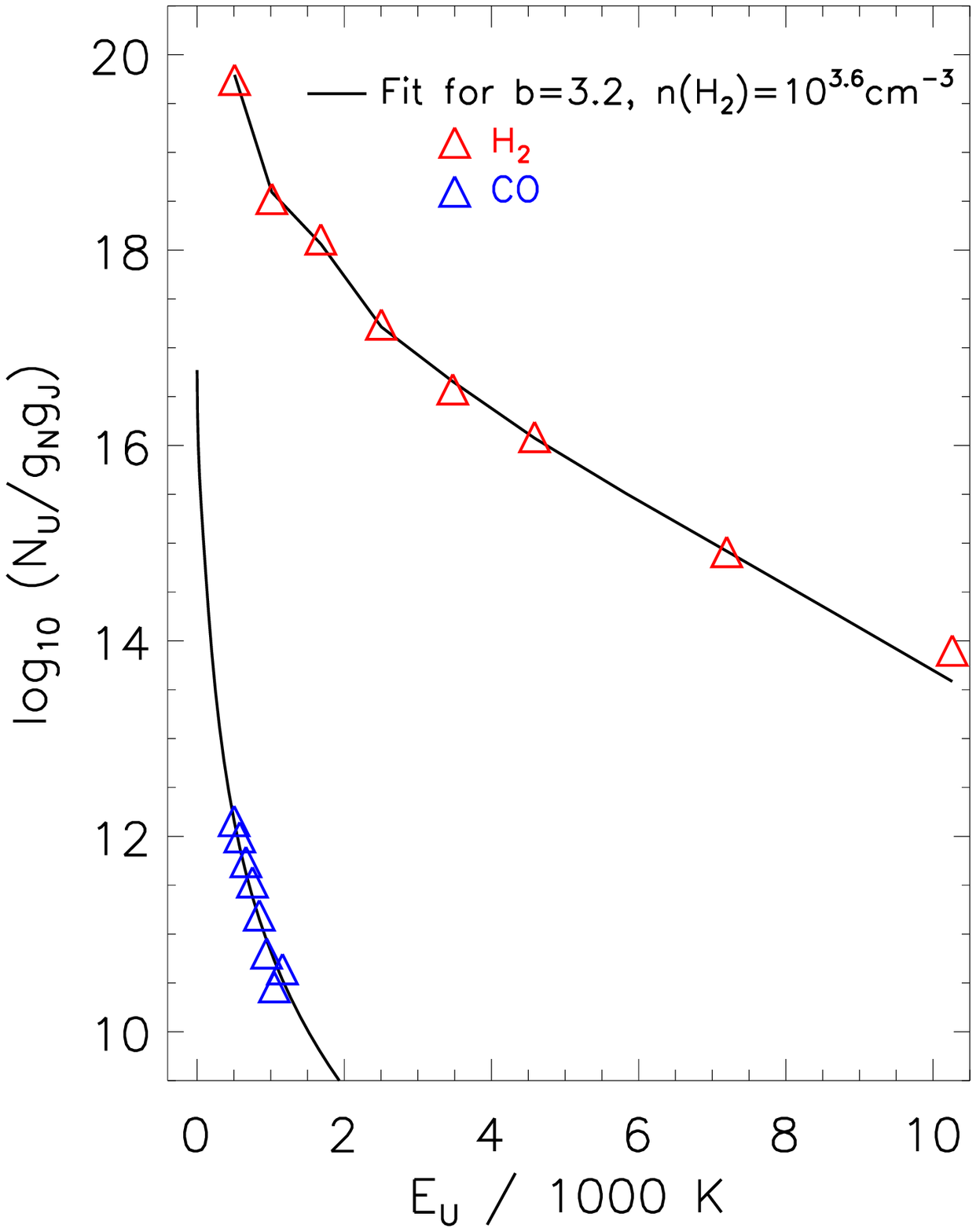}%

\includegraphics[width=8 cm, angle=0]{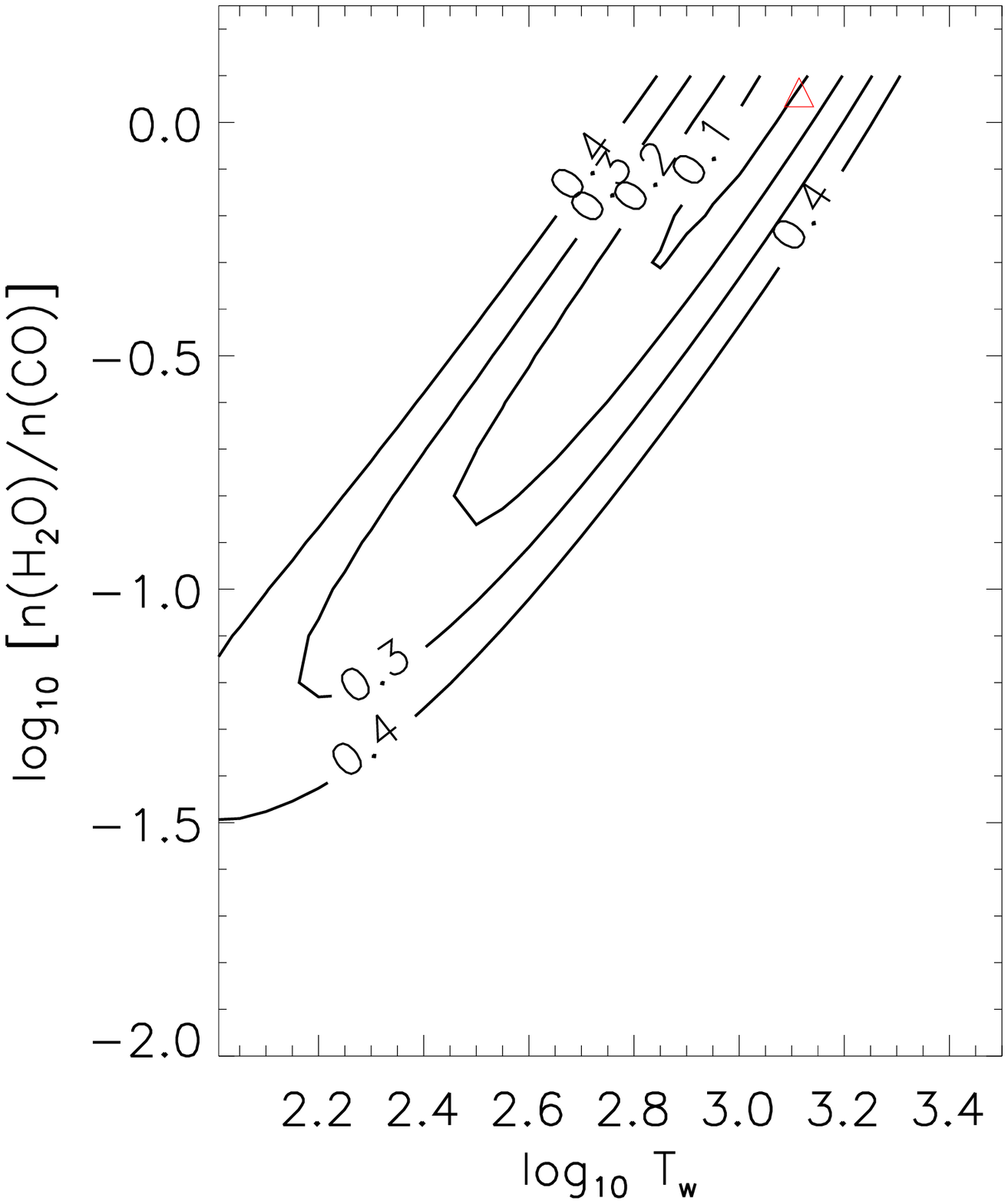}%
\includegraphics[width=8 cm, angle=0]{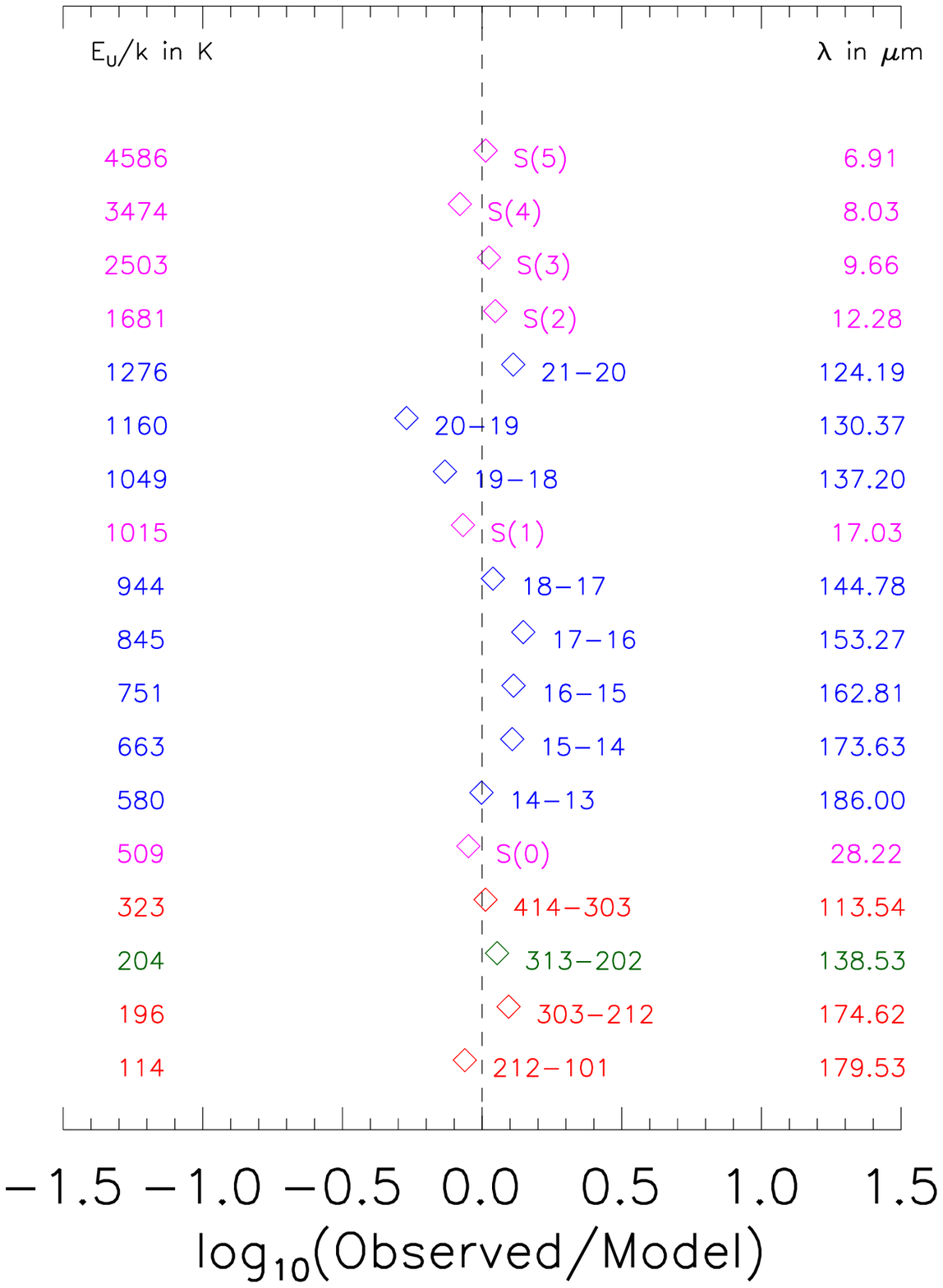}%                                        
\caption{Fit to the line fluxes observed from W28 (see text and Figure 12 caption for explanation)} 
\end{figure}

\end{document}